\documentclass[pra,reprint,superscriptaddress,twocolumn,preprintnumbers,amsmath,amssymb]{revtex4-1}
\usepackage{amssymb,latexsym}
\usepackage{amsbsy} % for \boldsymbol
\usepackage{amsmath,amsthm}
\usepackage{hyperref}
\usepackage{graphicx,color}
\usepackage{verbatim}

\newcommand\trace {\mathrm{Tr}}

\newcommand\jpvec {{\mathbf{j}_{\mathrm{p}}}}

\newcommand\nupvec {\boldsymbol{\nu}}

\newcommand\scp[2] {\langle {#1} | {#2} \rangle}

\begin{document}

\title{A local tensor that unifies kinetic energy density and vorticity dependent exchange-correlation functionals} 

\author{Sangita Sen}
\affiliation{Hylleraas Centre for Quantum Molecular Sciences, Department of Chemistry,  University of Oslo, P.O. Box 1033 Blindern, N-0315 Oslo, Norway}

\author{Erik I. Tellgren}
\email{erik.tellgren@kjemi.uio.no}
\affiliation{Hylleraas Centre for Quantum Molecular Sciences, Department of Chemistry,  University of Oslo, P.O. Box 1033 Blindern, N-0315 Oslo, Norway}

\begin{abstract}
We present a kinetic energy tensor that unifies a scalar kinetic energy density commonly used in meta-Generalized Gradient Approximation functionals and the vorticity density that appears in paramagnetic current-density-functional theory. Both types of functionals can thus be subsumed as special cases of a novel functional form that is naturally placed on the third rung of Jacob's ladder. Moreover, the kinetic energy tensor is related to the exchange hole curvature, is gauge invariant, and has very clearcut $N$-representability conditions. The latter conditions enable the definition of effective number of non-negligible orbitals. Whereas quantities such as the Electron Localization Function can discriminate effective one-orbital regions from other regions, the present kinetic energy tensor can discriminate between one-, two-, three-, and four-or-more orbital regions.
\end{abstract}

\maketitle 

\section{Introduction}

Density-functional theory has developed into several formal mathematical frameworks and numerous types of practical approximations. Current-density-functional theory (CDFT) is one generalization of the original formulation and provides a framework where all ground-state properties of an electronic system in external magnetic fields are determined by the density and paramagnetic current density~\cite{VIGNALE_PRL59_2360,CAPELLE_PRB65_113106}. In particular, the exchange-correlation energy is determined in this way. Gauge invariance requires that the exchange-correlation energy only depends on the density and the paramagnetic vorticity,
\begin{equation}
   \label{eqVORTICITYDEF}
   \nupvec(\mathbf{r}) = \nabla\times\frac{\jpvec(\mathbf{r})}{\rho(\mathbf{r})},
\end{equation}
where $\rho$ is the electron density and $\jpvec$ the gauge-dependent paramagnetic current density. Though the formal foundation of CDFT has been in place for three decades, the development of practical vorticity-dependent approximations is still in the early stages~\cite{VIGNALE_PRB37_2502,SKUDLARSKI_PRB48_8547,LEE_JCP103_10095,TAO_PRB74_193108,TELLGREN_JCP140_034101}. The available approximations are of the form
\begin{equation}
  \label{eqCDFTform}
  F_{\text{CDFT}}[\rho,\jpvec] = F_{\text{DFT}}[\rho] + \int f_{\text{CDFT}}(\rho(\mathbf{r}), \nabla\rho(\mathbf{r}), \nupvec(\mathbf{r})) \, d\mathbf{r},
\end{equation}
where $F_{\text{DFT}}[\rho]$ is any conventional exchange-correlation approximation and the second term is intended to correct for the current dependence. In what follows, we assume for simplicity that the vorticity-independent term is of the form $F_{\text{DFT}}[\rho] = \int f_{\mathrm{GGA}}(\rho,\nabla\rho) d\mathbf{r}$  so that it can be absorbed into the second term.

\begin{figure}
 \begin{center}
   \includegraphics[width=0.7\columnwidth]{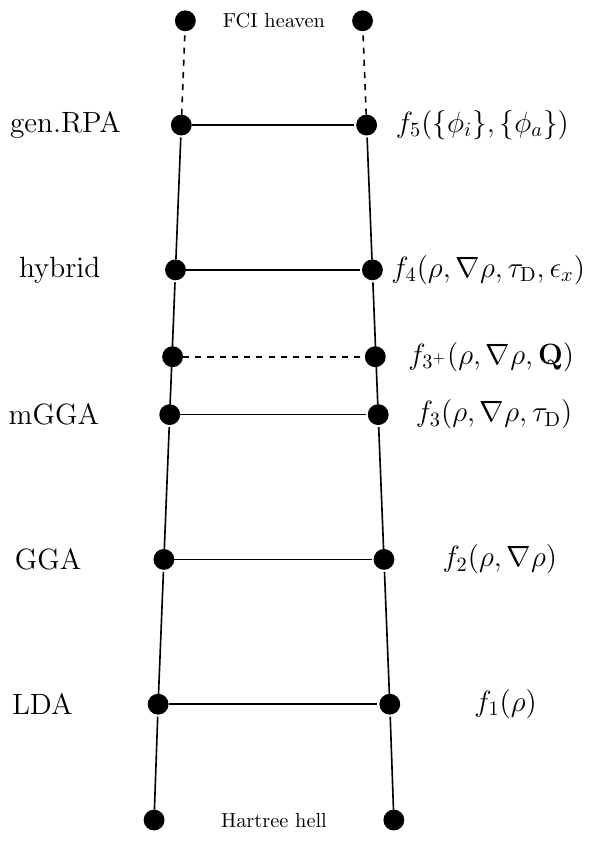}
 \end{center}
   \caption{The conventional Jacob's ladder classification of density-functional approximations. The third rung allows dependence on a kinetic energy density. The fourth rung adds dependence on the exact exchange density $\epsilon_x$ and the fifth allows generalized Random Phase Approximation-type models that include a dependence on both occupied and unoccupied orbitals. Vorticity-dependent functionals also belong on the third rung. Moreover, all rung 3 functionals can be subsumed as special cases of $\mathbf{Q}$-dependent functionals, here indicated as a slightly higher ``rung $3^+$''.}
  \label{figJACOB}
\end{figure}

The available density functional approximations are often classified based on their locality properties and placed on different rungs of Jacob's ladder~\cite{PERDEW_AIPCP577_1} illustrated in Fig.~\ref{figJACOB}. In contrast to the early stage of CDFT approximations, much recent effort has been directed at meta-Generalized Gradient Approximations (mGGA), which are placed on the third rung of Jacob's ladder. In the absence of magnetic fields, mGGAs take the form~\cite{BECKE_PRA39_3761,PERDEW_PRL82_2544}
\begin{equation}
  \label{eqMGGAform}
    F_{\text{mGGA}}[\rho,\tau_{\mathrm{D}}] = \int f_{\text{mGGA}}(\rho(\mathbf{r}), \nabla\rho(\mathbf{r}), \tau_{\mathrm{D}}(\mathbf{r})) \, d\mathbf{r},
\end{equation}
where $\tau_{\mathrm{D}} = \tau_{\mathrm{can}} - |\nabla\rho|^2/(8\rho)$ is conventionally taken to be related to the the everywhere positive, canonical kinetic energy density $\tau_{\mathrm{can}}$ of the Kohn--Sham system. (Less conventionally, it is possible to generalize DFT to include a kinetic energy density as an additional basic variable alongside the electron density~\cite{JANSEN_PRB43_12025,HIGUCHI_PRB69_035113,AYERS_PRA80_032510,AYERS_JCP126_144108}. In this case, $\tau_{\mathrm{D}}$ and $\rho$ would, at least formally if not practically, be densities of the interacting system that the noninteracting Kohn--Sham system should reproduce.) A few mGGAs also include a dependence on the density Laplacian $\nabla^2 \rho$, though for simplicity we suppress this from the notation. Many mGGA functionals rely on the fact that $\tau_{\mathrm{D}}$ is related to the exchange hole and that $\tau_{\mathrm{D}}$ can be used to construct measures of orbital overlap---isoorbital indicators---capable of detecting regions of space where only one orbital is non-negligible~\cite{BECKE_PRA39_3761,PERDEW_PRL82_2544,TAO_PRL91_146401,SUN_PRL111_106401}. In the presence of a magnetic field, the gauge dependence of $\tau_{\mathrm{can}}$ can no longer be resolved by restricting attention to real-valued wave functions. Two known gauge-invariant candidates exist to replace it. The physical kinetic energy density, which requires knowledge of the external magnetic vector potential in addition to the wave function, and Dobson's kinetic energy density~\cite{DOBSON_JCP94_4328,DOBSON_JCP98_8870}, which is determined by the wave function alone. The latter choice yields the gauge-corrected density
\begin{equation}
   \tau_{\mathrm{D}}(\mathbf{r}) = \tau_{\mathrm{can}}(\mathbf{r}) - \frac{|\nabla\rho(\mathbf{r})|^2}{8\rho(\mathbf{r})} - \frac{|\jpvec(\mathbf{r})|^2}{2\rho(\mathbf{r})}.
\end{equation}
We focus on this kinetic energy density as it is compatible with the CDFT framework and also retains the relationship to isoorbital indicators and models of the Hartree--Fock exchange hole in the presence of external magnetic fields~\cite{BECKE_CJC74_995,BATES_JCP137_164105,SAGVOLDEN_MP111_1295}. Recent work indicates that this type of functional is a promising practical alternative to the presently available vorticity dependent functionals~\cite{ZHU_PRA90_022504,FURNESS_JCTC11_4169,REIMANN_JCTC13_4089}.

The questions that prompted the present study are: {\it How are $\tau_{\mathrm{D}}$ and $\nupvec{}$ related? Can one be reconstructed from the other?}
A partial, formal answer can be given immediately. The pair $(\rho,\nupvec)$ determines $(\rho,\jpvec)$ to within a gauge.
Furthermore, CDFT admits a weak form of the Hohenberg--Kohn theorem with the implication that $(\rho,\jpvec)$ determines the ground-state wave function~\cite{CAPELLE_PRB65_113106}.
Hence, $(\rho,\nupvec)$ in principle determines all gauge invariant properties of the Kohn--Sham system too---in particular, $\tau_{\mathrm{D}}$ is determined. The formal mapping from $(\rho,\tau_{\mathrm{D}})$ to $(\rho,\nupvec)$ is less clear. On a more practical level, we will demonstrate below that $\tau_{\mathrm{D}}$ and $\nupvec$ are essentially independent components of a kinetic energy-like tensor. We will also establish strong $N$-representability conditions on the kinetic tensor, which enable new measures of orbital overlap that are more powerful than existing isoorbital indicators.

\section{Vorticity and Dobson's kinetic energy density as tensor components}

Any wave function gives rise to a one-particle reduced density matrix, which can be diagonalized to obtain natural orbitals $\phi_l$ and occupation numbers $n_l$. In the present setting, the natural orbitals are the Kohn--Sham orbitals and it is common to take the occupation numbers to be integers $n_l \in \{0,2\}$ corresponding to unoccupied and doubly occupied spatial orbitals, respectively. Since all equations below refer only to occupied orbitals, we simplify the notation by absorbing occupation numbers into the normalization condition, $\scp{\phi_l}{\phi_l} = n_l$.  The basic CDFT densities are then given by
\begin{align}
   \label{eqRhoFromOrb}
   \rho(\mathbf{r}) & = \sum_l |\phi_l(\mathbf{r})|^2, \\
   \jpvec(\mathbf{r}) & = \frac{1}{2} \sum_l \phi_l(\mathbf{r})^* \, \mathbf{p} \, \phi_l(\mathbf{r}) + \mathrm{c.c.},
\end{align}
where $\mathbf{p} = -i\nabla$ is the canonical momentum operator. The physical current density is given by $\mathbf{j} = \jpvec + \rho\mathbf{A}$, where $\mathbf{A}$ is the external magnetic vector potential, but it is the gauge dependent $\jpvec$ that is a basic variable in CDFT. It is useful to also introduce the complex current density
\begin{equation}
  \boldsymbol{\kappa}(\mathbf{r}) = \sum_l \phi_l(\mathbf{r})^* \, \mathbf{p} \, \phi_l(\mathbf{r}) = \jpvec(\mathbf{r}) - \frac{i}{2} \nabla \rho(\mathbf{r}).
\end{equation}
The paramagnetic vorticity in Eq.~\eqref{eqVORTICITYDEF} can be written
\begin{equation}
   \label{eqVORTICITYALT}
   \begin{split}
     \nupvec(\mathbf{r})  & = \frac{\nabla\times\jpvec(\mathbf{r})}{\rho(\mathbf{r})} + \frac{\jpvec(\mathbf{r})\times\nabla\rho(\mathbf{r})}{\rho(\mathbf{r})^2}
                       \\
         & = \frac{\nabla\times\boldsymbol{\kappa}(\mathbf{r})}{\rho(\mathbf{r})} - i \, \frac{\boldsymbol{\kappa}(\mathbf{r})\times\boldsymbol{\kappa}(\mathbf{r})^*}{\rho(\mathbf{r})^2}.
   \end{split}
\end{equation}
Using the identity $\nabla\times\nabla\phi_l = 0$, one finds that $\nabla\times\jpvec = i\sum_l (\nabla\phi_l)\times \nabla\phi_l^*$ and none of the contributions to $\nupvec$ contains second-derivatives of the orbitals. From a computational point of view, the calculation of $\nupvec$ is thus comparable to the calculation of everywhere positive kinetic energy densities. Hence, the vorticity-dependent functionals can be placed alongside common mGGAs on the third rung of Jacob's ladder.

The canonical kinetic energy tensor is defined as
\begin{equation}
    \tau_{ab}(\mathbf{r}) = \frac{1}{2} \sum_l \big( p_a \phi_l(\mathbf{r}) \big) \, \big( p_b \phi_l(\mathbf{r}) \big)^*,
\end{equation}
where $a,b\in\{1,2,3\}$ run over the three Cartesian directions. This tensor is gauge dependent. Its trace, $\tau_{\mathrm{can}} = \tau_{11} + \tau_{22} + \tau_{33}$, is the usual canonical kinetic energy density---a scalar quantity that is non-negative everywhere in space, $\tau_{\mathrm{can}}(\mathbf{r}) \geq 0$. A gauge invariant kinetic energy tensor is readily obtained through the minimal substitution $\mathbf{p} \to \boldsymbol{\pi}_{\mathbf{A}} = \mathbf{p} + \mathbf{A}$. However, the explicit dependence on the external potential $\mathbf{A}$ makes this unsuitable for a CDFT setting. Instead, we define the {\it intrinsic kinetic energy tensor} as
\begin{equation}
   \label{eqIntrinsicKinDef}
   Q_{ab}(\mathbf{r}) = \tau_{ab}(\mathbf{r}) - \frac{\kappa_a(\mathbf{r}) \, \kappa_b(\mathbf{r})^*}{2 \rho(\mathbf{r})}.
\end{equation}
This tensor is gauge invariant. In fact, the ratio $\mathbf{Q}/\rho$ is even invariant with respect local scaling $\phi_l(\mathbf{r}) \mapsto \phi'_l(\mathbf{r}) = \phi_l(\mathbf{r}) \, \Omega(\mathbf{r})$ of orbitals by a complex function $\Omega$. The special case $|\Omega(\mathbf{r})| \equiv 1$ gives a gauge transformation.

The trace of $\mathbf{Q}$ is equal (to within a von Weizs\"acker term) to the kinetic energy density introduced by Dobson~\cite{DOBSON_JCP94_4328},
\begin{equation}
     \label{eqTauDFromQ}
   \tau_{\mathrm{D}} = Q_{11} + Q_{22} + Q_{33} = \tau_{\mathrm{can}} - \frac{|\jpvec|^2}{2\rho} - \frac{|\nabla\rho|^2}{8\rho}.
\end{equation}
Moreover, the off-diagonal elements of $\mathbf{Q}$ encode the vorticity. To establish this, we write a particular off-diagonal element of $\boldsymbol{\tau}$ as
\begin{equation}
   \tau_{12} = \frac{1}{2} \sum_l (\nabla_1 \phi_l) \, \nabla_2 \phi_l^*.
\end{equation}
Comparison with the third Cartesian component of $\nabla\times\jpvec$,
\begin{equation}
   [\nabla\times\jpvec]_3 = i \sum_l \big( (\nabla_1 \phi_l) \, \nabla_2 \phi_l^* - (\nabla_2 \phi_l) \, \nabla_1 \phi_l^* \big),
\end{equation}
then directly yields $[\nabla\times\jpvec]_3 = 2i(\tau_{12} - \tau_{21})$. In general, with $\epsilon_{cab}$ denoting the Levi-Civita tensor and implicit summation over $a,b$,
\begin{equation}
   [\nabla\times\jpvec]_c = 2i \, \epsilon_{cab} \tau_{ab}.
\end{equation}
Hence, the first term in Eq.~\eqref{eqVORTICITYALT} is encoded in the anti-symmetric, imaginary part of the tensor $\boldsymbol{\tau}/\rho$. Similarly, we find that the second term in Eq.~\eqref{eqVORTICITYALT} is encoded in the second term of Eq.~\eqref{eqIntrinsicKinDef}. We have $\boldsymbol{\kappa}\times\boldsymbol{\kappa}^* = i \, \jpvec \times \nabla\rho$ as well as
\begin{equation}
   [\jpvec\times\nabla\rho]_c = -i \, \epsilon_{cab} \kappa_a \kappa_b^*.
\end{equation}
By combining the above results, it follows that the anti-symmetric part of the intrinsic kinetic energy tensor $\mathbf{Q}$ encodes the vorticity,
\begin{equation}
   \label{eqNuFromQ}
   \nu_c = 2i \frac{\epsilon_{cab} Q_{ab}}{\rho}.
\end{equation}

To summarize, $\mathbf{Q}$ is gauge invariant and it encodes both Dobson's scalar kinetic energy density in its trace (see Eq.~\eqref{eqTauDFromQ}) and the vorticity in its anti-symmetric part (see Eq.~\eqref{eqNuFromQ}). Hence, the form
\begin{equation}
  F[\rho,\mathbf{Q}] = \int f(\rho(\mathbf{r}),\nabla\rho(\mathbf{r}),\mathbf{Q}(\mathbf{r})) \, d\mathbf{r}
\end{equation}
contains both the vorticity-dependent form in Eq.~\eqref{eqCDFTform} and the mGGA form in Eq.~\eqref{eqMGGAform} as special cases:
\begin{align}
    F_{\text{CDFT}}[\rho,\nupvec] & = \int f_{\text{CDFT}}(\rho, \nabla\rho, i(\mathbf{Q}-\mathbf{Q}^T)/\rho) \, d\mathbf{r},
       \\
   F_{\text{mGGA}}[\rho,\tau_{\mathrm{D}}] & = \int f_{\text{mGGA}}(\rho, \nabla\rho, \trace(\mathbf{Q})) \, d\mathbf{r}.
 \label{eqMGGAForm}
\end{align}
This generalization is one of our main results. Besides unifying the two classes of density functional approximations, it also provides a new framework for incorporating vorticity-dependence into isoorbital indicators and mGGA functionals. This problem has been raised but not solved in the literature~\cite{TAO_PRL95_196403,SAGVOLDEN_MP111_1295}.

\section{Exchange hole curvature}

The tensor $\mathbf{Q}$ was introduced above based on considerations of gauge invariance. The symmetric part of $\mathbf{Q}$ could have also been obtained as the Hessian of the exchange hole. A closed-shell Slater determinant made up from orbitals $\phi_1,\ldots,\phi_K$ gives rise to a one-particle reduced density matrix
\begin{equation}
    D(\mathbf{r},\mathbf{s}) = \sum_l \phi_l(\mathbf{r}) \phi_l(\mathbf{s})^*
\end{equation}
and the pair density
\begin{equation}
 \begin{split}
    \Gamma(\mathbf{r},\mathbf{s}) & = D(\mathbf{r},\mathbf{r}) D(\mathbf{s},\mathbf{s}) - \frac{1}{2} D(\mathbf{r},\mathbf{s}) D(\mathbf{s},\mathbf{r}) 
            \\
      &= \rho(\mathbf{r}) \rho(\mathbf{s}) - \Gamma_{\mathrm{X}}(\mathbf{r},\mathbf{s}),
 \end{split}
\end{equation}
where we identify the electron density as the diagonal of the density matrix, $\rho(\mathbf{r}) = D(\mathbf{r},\mathbf{r})$, and denote the second, exchange-like term by $\Gamma_{\mathrm{X}}(\mathbf{r},\mathbf{s})$. A simple calculation now verifies that
\begin{align}
   \frac{\partial^2 \Gamma_{\mathrm{X}}(\mathbf{r},\mathbf{s})}{\partial s_a \partial s_b} \Big|_{\mathbf{s}=\mathbf{r}} = \frac{\kappa_a \kappa_b^* + \kappa_a^* \kappa_b}{2} - \rho (t^{\mathrm{Lap}}_{ab} + t^{\mathrm{Lap}*}_{ab}),
\end{align}
where $t^{\mathrm{Lap}}_{ab} = -\tfrac{1}{2} \sum_l \phi_l(\mathbf{r})^* \nabla_a \nabla_b \phi_l(\mathbf{r})$ is a Laplacian form of the kinetic energy tensor. Inserting the identity $t^{\mathrm{Lap}}_{ab} + t^{\mathrm{Lap}*}_{ab} = \tau_{ab} + \tau_{ba} - \tfrac{1}{2} \nabla_a \nabla_b \rho$ yields
\begin{equation}
   \label{eqQExchHessian}
 \begin{split}
   \frac{\partial^2 \Gamma_{\mathrm{X}}(\mathbf{r},\mathbf{s})}{\partial s_a \partial s_b} \Big|_{\mathbf{s}=\mathbf{r}} & = \frac{\kappa_a \kappa_b^* + \kappa_a^* \kappa_b}{2} - \rho (\tau_{ab} + \tau_{ba}) + \frac{\rho \nabla_a \nabla_b \rho}{2}
           \\
     & = -\rho (Q_{ab} + Q_{ba}) + \frac{\rho \nabla_a \nabla_b \rho}{2}.
 \end{split}
\end{equation}
Hence, the real, symmetric part of $\mathbf{Q}$ is related to the exchange hole Hessian. The trace of the Hessian recovers the well-known spherically averaged exchange hole curvature, often considered in spin-resolved form~\cite{BECKE_IJQC23_1915,DOBSON_JCP94_4328}.

\section{$N$-representability of the intrinsic kinetic energy tensor}

A prescribed value for a quantity is said to be \emph{$N$-representable} if there exists an $N$-electron wave function or, more generally, a mixed state that reproduces this value. It is well-known that both the scalar density $\tau_{\mathrm{D}}$ and the vorticity field $\nupvec$ vanish identically in regions of space where only a single orbital gives a non-negligible contribution. In general, essentially all pairs $(\rho,\jpvec)$, subject only to very mild mathematical regularity conditions, are mixed state $N$-representable~\cite{TELLGREN_PRA89_012515}. The same is true for Slater-determinantal (noninteracting) $N$-representability with four or more orbitals~\cite{LIEB_PRA88_032516}. The conditions for two- and three-orbital $N$-representability of $(\rho,\jpvec)$ are presently open questions. A gap in the literature is the lack of $N$-representability conditions for related tensor quantities. Below, we address this gap and provide simple necessary conditions for $N$-representability of $\boldsymbol{\tau}$ and $\mathbf{Q}$. It is plausible that these are also sufficient conditions, though a rigorous proof is lacking.

\subsection{Rank of $\mathbf{Q}$ as a bound on the number of orbitals}

In what follows, we shall take the number of Kohn--Sham (or natural) orbitals with non-zero occupation to be $K$ and the number of electrons to be $N = \sum_{l=1}^K n_l = \sum_{l=1}^K \scp{\phi_l}{\phi_l}$. For a closed-shell Kohn--Sham system, $N = 2K$. Writing
\begin{equation}
   \boldsymbol{\tau} = \frac{1}{2} \sum_{l=1}^K (\mathbf{p} \phi_l) (\mathbf{p} \phi_l)^{\dagger},
\end{equation}
we note that, at any point in space, $\boldsymbol{\tau}$ is a sum of $K$ outer products. It follows that the $3\times 3$ kinetic energy tensor $\boldsymbol{\tau}$ is positive semi-definite and that its rank cannot exceed the number of terms,
\begin{equation}
   \mathrm{rank}( \boldsymbol{\tau} ) \leq \min(K,3).
\end{equation}
Since $\boldsymbol{\tau} = \boldsymbol{\tau}^{\dagger}$ is hermitian, $\mathrm{rank}(\boldsymbol{\tau})$ is equivalent to the number of non-zero eigenvalues. This is a useful $N$-representability result, since it provides a clear-cut restriction on which tensors $\boldsymbol{\tau}$ can be obtained from $K=1$ and $K=2$ orbital systems.

Turning to the intrinsic kinetic energy tensor $\mathbf{Q}$ it can be verified that it vanishes identically everywhere in space for single orbital ($K=1$) systems. This unifies the known conditions $\tau_{\mathrm{D}} = \trace(\mathbf{Q}) = 0$ and $\mathbf{Q} - \mathbf{Q}^{T} \leftrightarrow \nupvec = \mathbf{0}$ into the stronger condition $\mathbf{Q} = \mathbf{0}$. For arbitrary $K$, we write
\begin{equation}
  Q_{ab} = \frac{1}{2\rho} \sum_{l,j=1}^K (p_a \phi_l) \, (\rho \delta_{lj} - \phi_l^* \phi_j) \, (p_b \phi_j)^*.
\end{equation}
Defining the $3\times K$ matrix $P_{al} = p_a \phi_l$ and arranging the complex-conjugated orbitals into a column vector $\bar{\boldsymbol{\Phi}} = (\phi_1^*,\ldots,\phi_K^*)^T$ now leads to the matrix form
\begin{equation}
   \mathbf{Q} = \frac{1}{2\rho} \, \mathbf{P} \,  (\bar{\boldsymbol{\Phi}}^{\dagger} \bar{\boldsymbol{\Phi}} \mathbf{I} - \bar{\boldsymbol{\Phi}} \bar{\boldsymbol{\Phi}}^{\dagger}) \, \mathbf{P}^{\dagger},
\end{equation}
where $\mathbf{I}$ is the $K\times K$ identity matrix and we have used Eq.~\eqref{eqRhoFromOrb} in the form $\rho = \bar{\boldsymbol{\Phi}}^{\dagger} \bar{\boldsymbol{\Phi}}$. Introducing the $K\times K$ matrix $\mathcal{M} = \bar{\boldsymbol{\Phi}}^{\dagger} \bar{\boldsymbol{\Phi}} \mathbf{I} - \bar{\boldsymbol{\Phi}} \bar{\boldsymbol{\Phi}}^{\dagger}$, we note that $\mathcal{M}/\rho$ is a projector onto the orthogonal complement of $\bar{\boldsymbol{\Phi}}$. Additionally,
\begin{align}
  \bar{\boldsymbol{\Phi}}^{\dagger} \mathcal{M} \bar{\boldsymbol{\Phi}} & = 0,
            \\
  \boldsymbol{\Xi}^{\dagger} \mathcal{M} \boldsymbol{\Xi} & =  \bar{\boldsymbol{\Phi}}^{\dagger} \bar{\boldsymbol{\Phi}} \, \boldsymbol{\Xi}^{\dagger} \boldsymbol{\Xi} > 0, \quad \text{for all} \ \boldsymbol{\Xi} \perp \bar{\boldsymbol{\Phi}}.
\end{align}
It follows that $\mathcal{M}$ is positive definite on the space of vectors orthogonal to $\bar{\boldsymbol{\Phi}}$. Hence, $\mathcal{M}$ is a rank $K-1$ matrix. Noting that $\mathrm{rank}(\mathbf{P}) \leq \min(K,3)$, we obtain our second main result:
\begin{equation}
   \mathrm{rank}( \mathbf{Q} ) = \mathrm{rank}( \mathbf{P} \mathcal{M} \mathbf{P}^{\dagger} ) \leq \min(K-1,3).
\end{equation}
and, as a consequence of the positive semidefiniteness of $\mathcal{M} \geq 0$,
\begin{equation}
    \mathbf{Q} = \boldsymbol{\tau} - \frac{\boldsymbol{\kappa} \boldsymbol{\kappa}^{\dagger}}{2\rho} \geq 0.
\end{equation}

The above result is a powerful $N$-representability condition on $\mathbf{Q}$. Moreover, the positive semidefiniteness gives a tensor generalization of the standard von Weizs\"acker lower bound on the kinetic energy density $\tau_{\mathrm{can}} \geq |\boldsymbol{\kappa}|^2/2\rho$~\cite{BATES_JCP137_164105}, with equality in the single-orbital ($K=1$) case.
In order to have full rank, at least $K=4$ orbitals are required. Moreover, though exceptions are possible, the typical case is $\mathrm{rank}(\mathbf{Q}) = K-1$ when $K \leq 4$. Besides the advantage of gauge invariance, the intrinsic kinetic energy tensor $\mathbf{Q}$ is thus more informative than the canonical tensor. The former can discriminate between $K=3$ and $K=4$, whereas the latter typically has full rank in both these cases.

\subsection{Upper bounds on vorticity}

The imaginary, anti-symmetric part of $\mathbf{Q}$,
\begin{equation}
   \boldsymbol{\Omega} = \frac{1}{2} (\mathbf{Q} - \mathbf{Q}^T) = \frac{1}{2} (\mathbf{Q} - \mathbf{Q}^*),
\end{equation}
directly encodes the vorticity vector in matrix form. Eq.~\eqref{eqNuFromQ} can be rewritten as
\begin{equation}
  \boldsymbol{\Omega} = 
        \frac{i \, \rho}{4} \begin{pmatrix}
               0 & -\nu_3 & \nu_2 \\
             \nu_3  & 0 & -\nu_1 \\
            -\nu_2 & \nu_1 & 0
           \end{pmatrix}.
\end{equation}
This encoding of an axial vector within an anti-symmetric matrix can be compared to how the magnetic field appears in the electromagnetic field tensor. By choosing the local coordinate axes so that $\nu_1 = \nu_2 = 0$, it is seen that the eigenvalues of $\boldsymbol{\Omega}$ are $0$, $-\rho|\nupvec|/4$, and $\rho|\nupvec|/4$. For any matrix norm with the property $\|\mathbf{M}\| = \|\mathbf{M}^*\|$ it now follows from the triangle inequality that
\begin{equation}
   \|\boldsymbol{\Omega}\| = \frac{1}{2} \|\mathbf{Q} - \mathbf{Q}^*\| \leq  \|\mathbf{Q}\|.
\end{equation}
Specific choices of matrix norms yields upper bounds on the vorticity. For example, the Schatten norm is defined as
\begin{equation}
    \|\mathbf{M}\|_{\alpha} = \Big( \sum_l |\mu_l|^{\alpha} \Big)^{1/\alpha}, \quad \alpha \geq 1,
\end{equation}
where $\mu_l$ is the $l$:th eigenvalue (or, more generally, singular value) of $\mathbf{M}$. Because $\mathbf{Q}$ is hermitian and positive semidefinite, we have $\|\mathbf{Q}\|_{\alpha}^{\alpha} = \trace(\mathbf{Q}^{\alpha})$. Choosing the Schatten norm in the above inequality yields
\begin{equation}
   \frac{2^{1/\alpha} \, \rho |\nupvec|}{4} = \|\boldsymbol{\Omega}\|_{\alpha} \leq  \|\mathbf{Q}\|_{\alpha}
\end{equation}
Equivalently, in terms of the three eigenvalues $q_1,q_2,q_3 \geq 0$ of $\mathbf{Q}$,
\begin{equation}
   \frac{\rho |\nupvec|}{4} \leq \Big( \frac{q_1^{\alpha} + q_2^{\alpha} + q_3^{\alpha}}{2} \Big)^{1/\alpha}.
\end{equation}
The particular choice $\alpha=1$ yields the trace norm and an upper bound in terms of Dobson's kinetic energy density,
\begin{equation}
   \label{eqNuTauDBound}
   \frac{\rho |\nupvec|}{2} = \|\boldsymbol{\Omega}\|_1 \leq  \|\mathbf{Q}\|_1 = \tau_{\mathrm{D}}.
\end{equation}

\begin{figure}
  \includegraphics[width=\columnwidth]{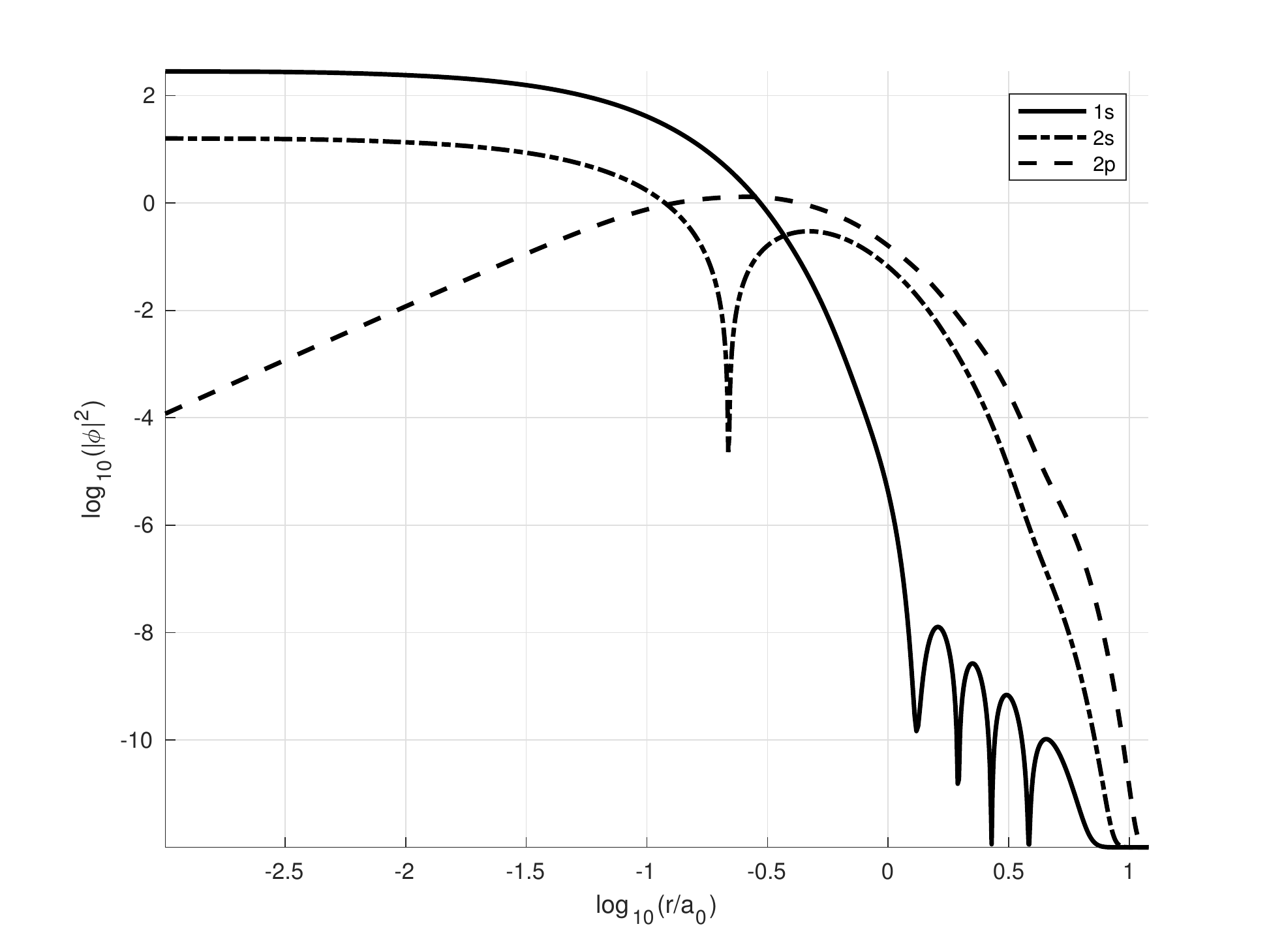}  \\
  \includegraphics[width=\columnwidth]{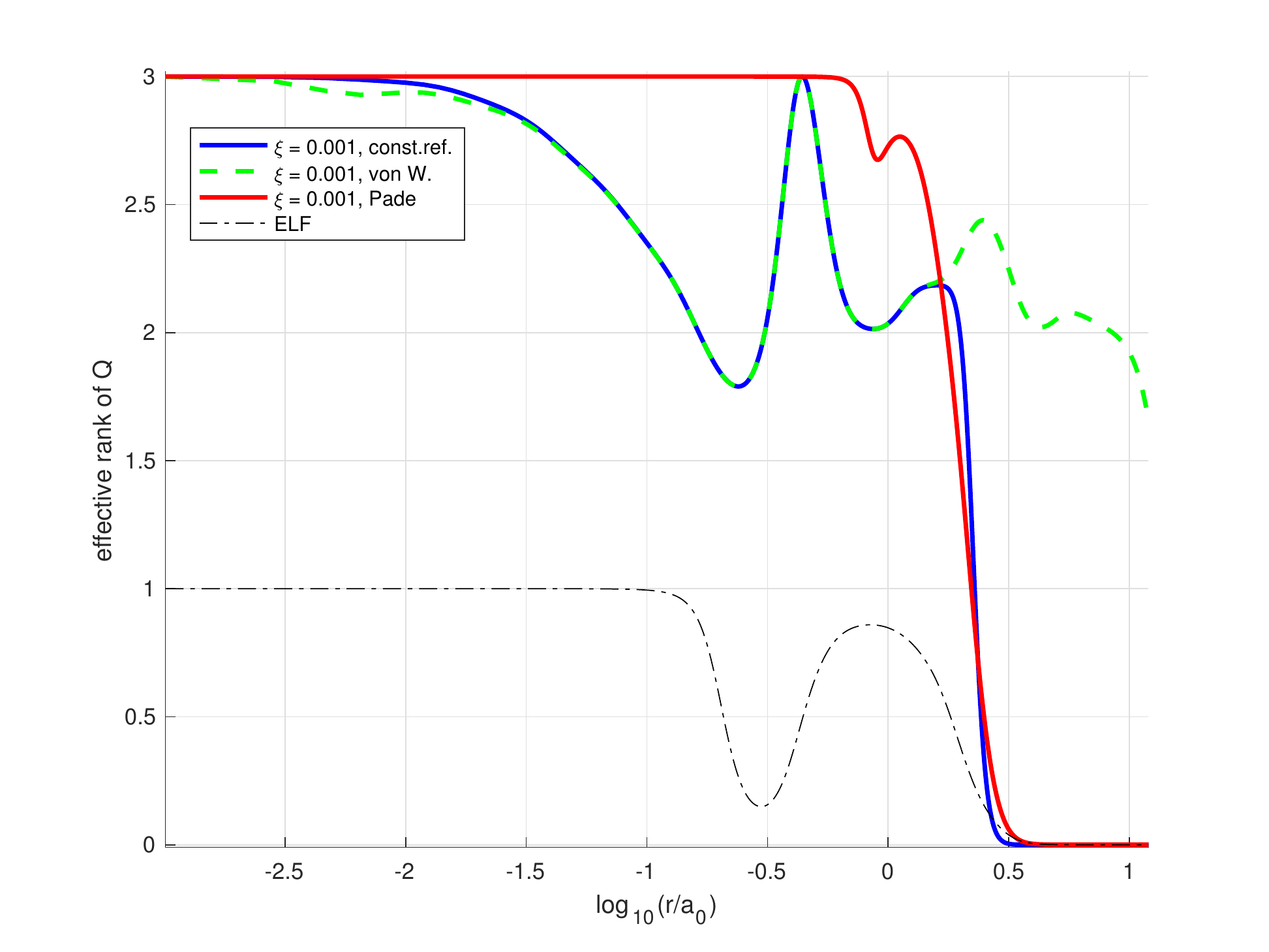}
  \caption{Top panel: The occupied orbitals in the neon atom. The finite local minum in the 2s curve is a numerical artifact---none of the sampling points exactly coincides with the nodal surface. Bottom: Effective rank estimates, obtained with $\xi = 10^{-3}$, along a radial line away from the atom. The solid blue line shows results for a position-independent reference $\tau_{\mathrm{ref}} \equiv 1$, the dashed green curve for a von Weizs\"acker reference energy $\tau_{\mathrm{ref}} = \frac{1}{2} |\nabla \sqrt{\rho}|^2$, and the solid red curve shows the Pad\'e approximation. The black dot-dashed line shows the commonly used Electron Localization Function (ELF).}
  \label{figLineNe}
\end{figure}

\begin{figure}
  \includegraphics[width=\columnwidth]{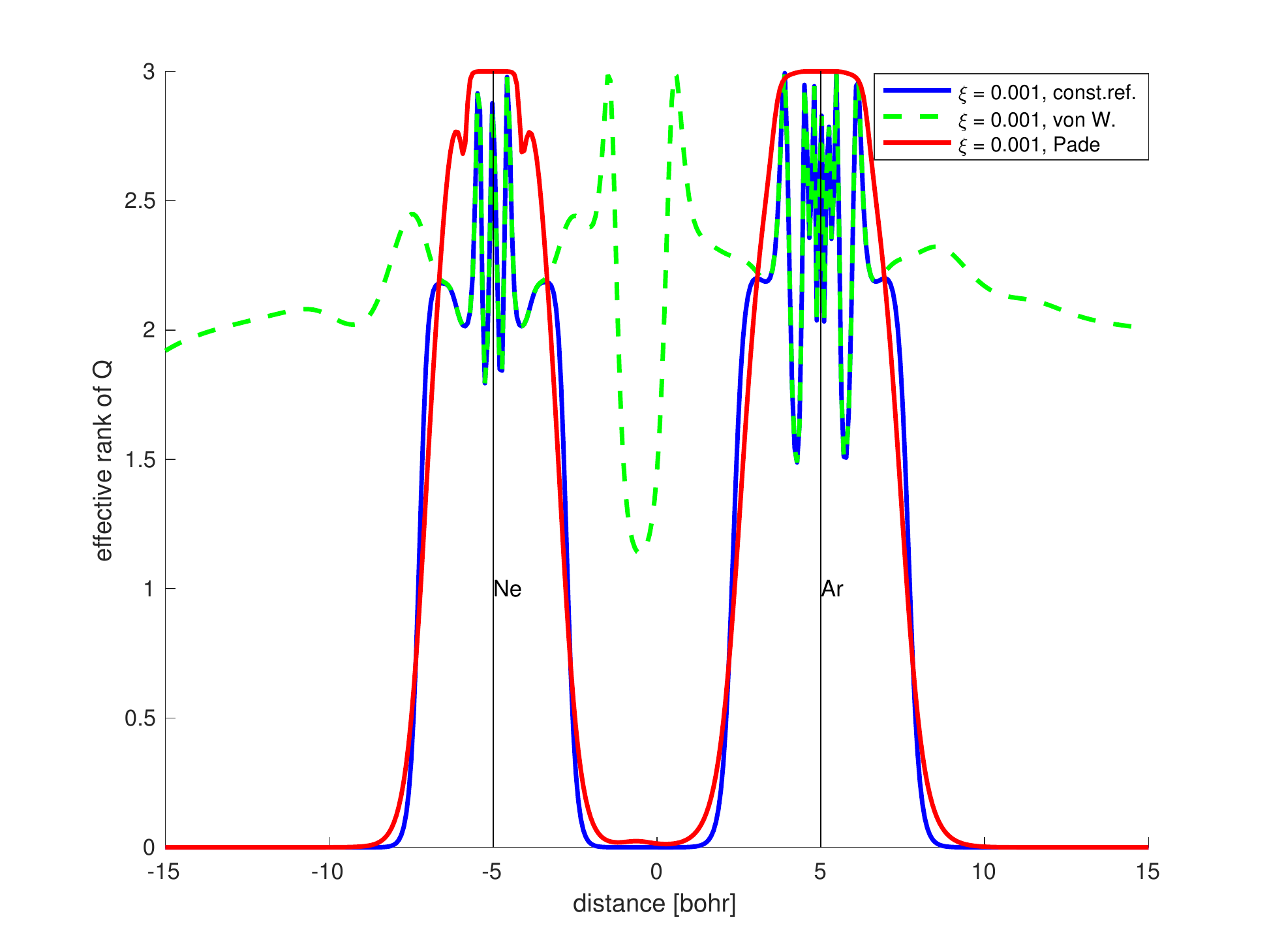} \\
  \includegraphics[width=\columnwidth]{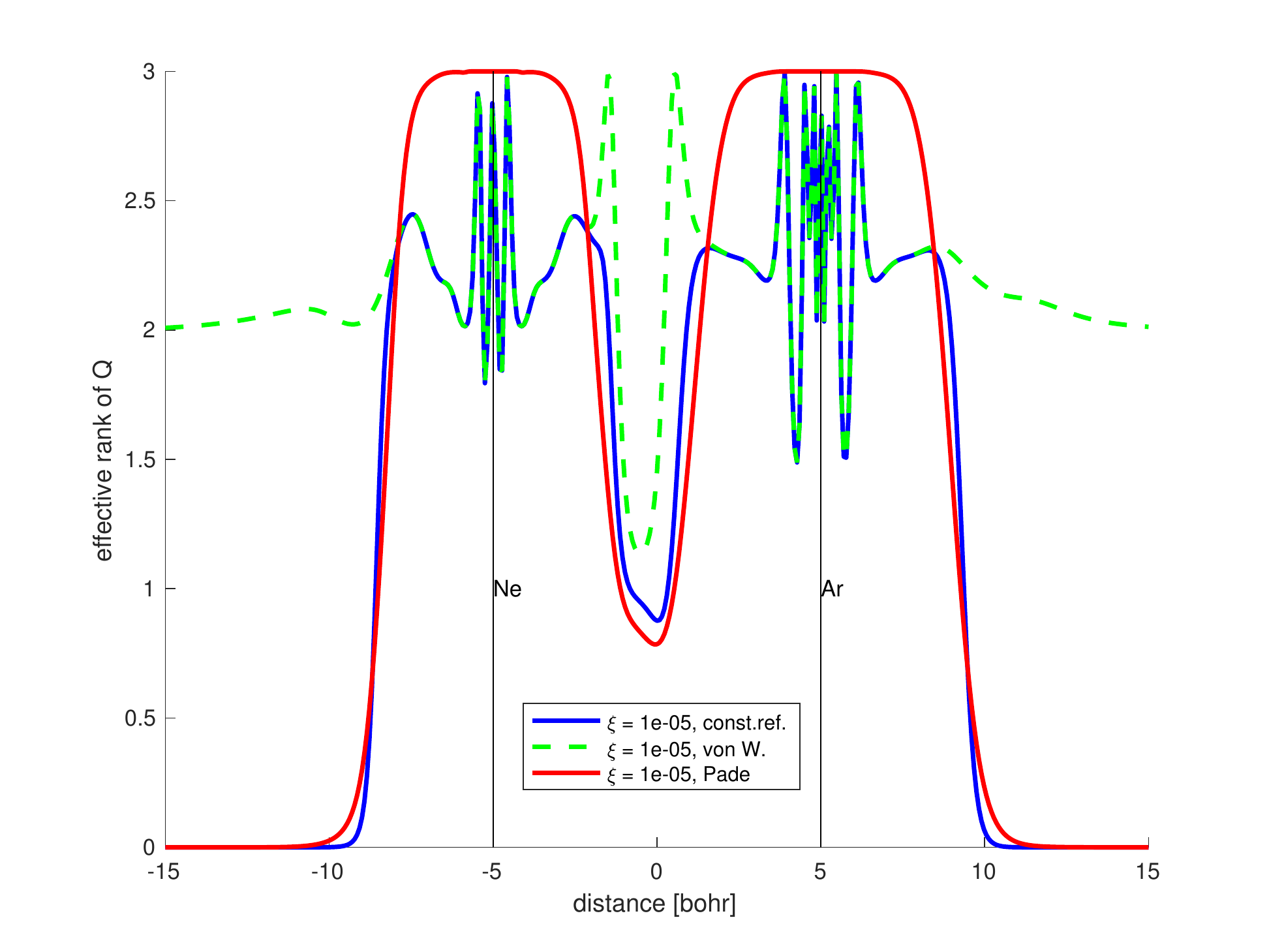}
  \caption{Effective rank estimates at different locations on a line through a neon and an argon atom, separated by 10~bohr. The top and bottom panels show results for $\xi = 10^{-3}$ and $\xi = 10^{-5}$, respectively. The solid blue line shows results for a position-independent reference $\tau_{\mathrm{ref}} \equiv 1$, the dashed green curve for a von Weizs\"acker reference energy $\tau_{\mathrm{ref}} = \frac{1}{2} |\nabla \sqrt{\rho}|^2$, and the solid red curve shows the Pad\'e approximation.}
  \label{figLineNeAr}
\end{figure}

\begin{figure}
  \includegraphics[width=\columnwidth]{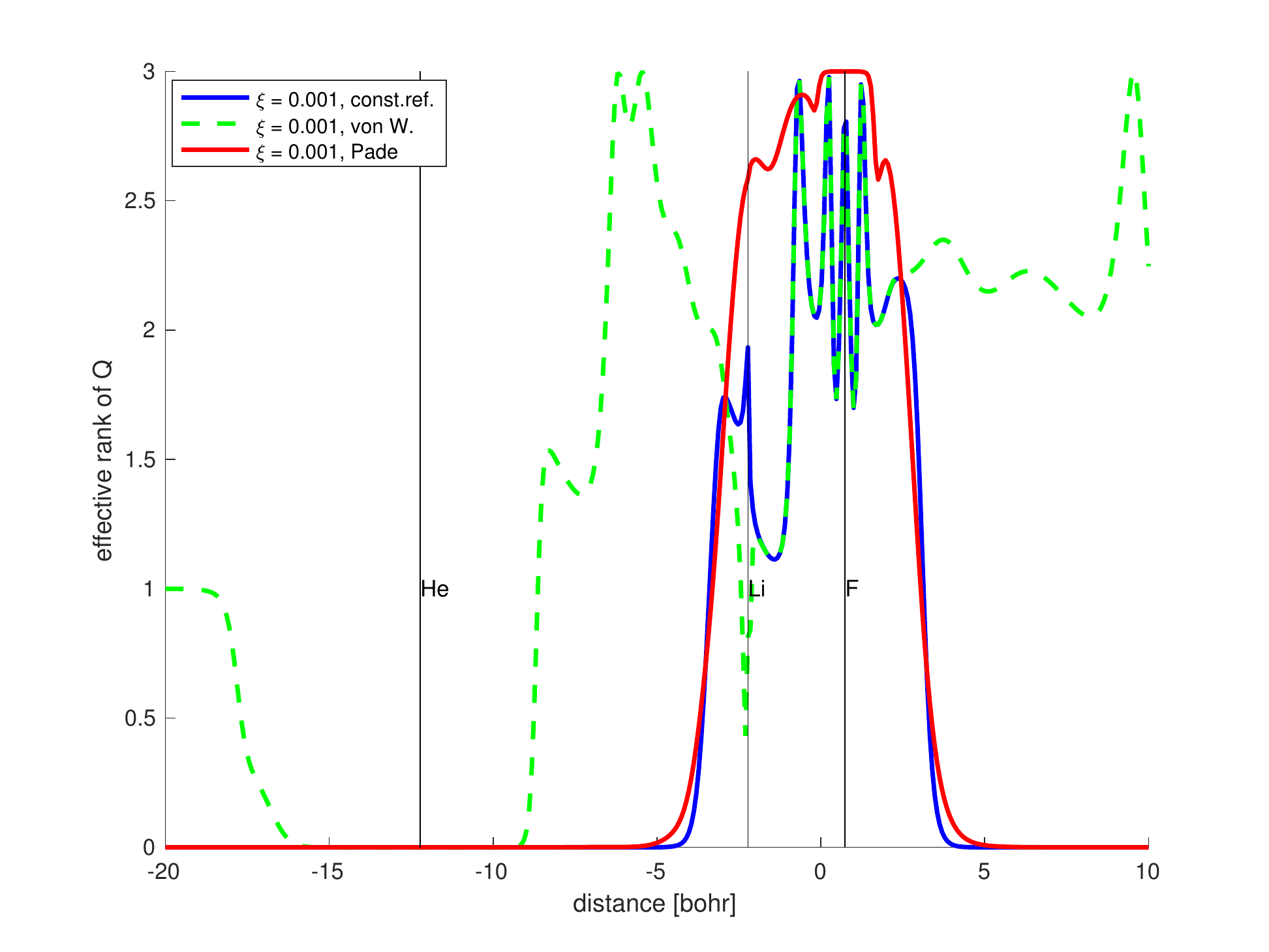} \\
  \includegraphics[width=\columnwidth]{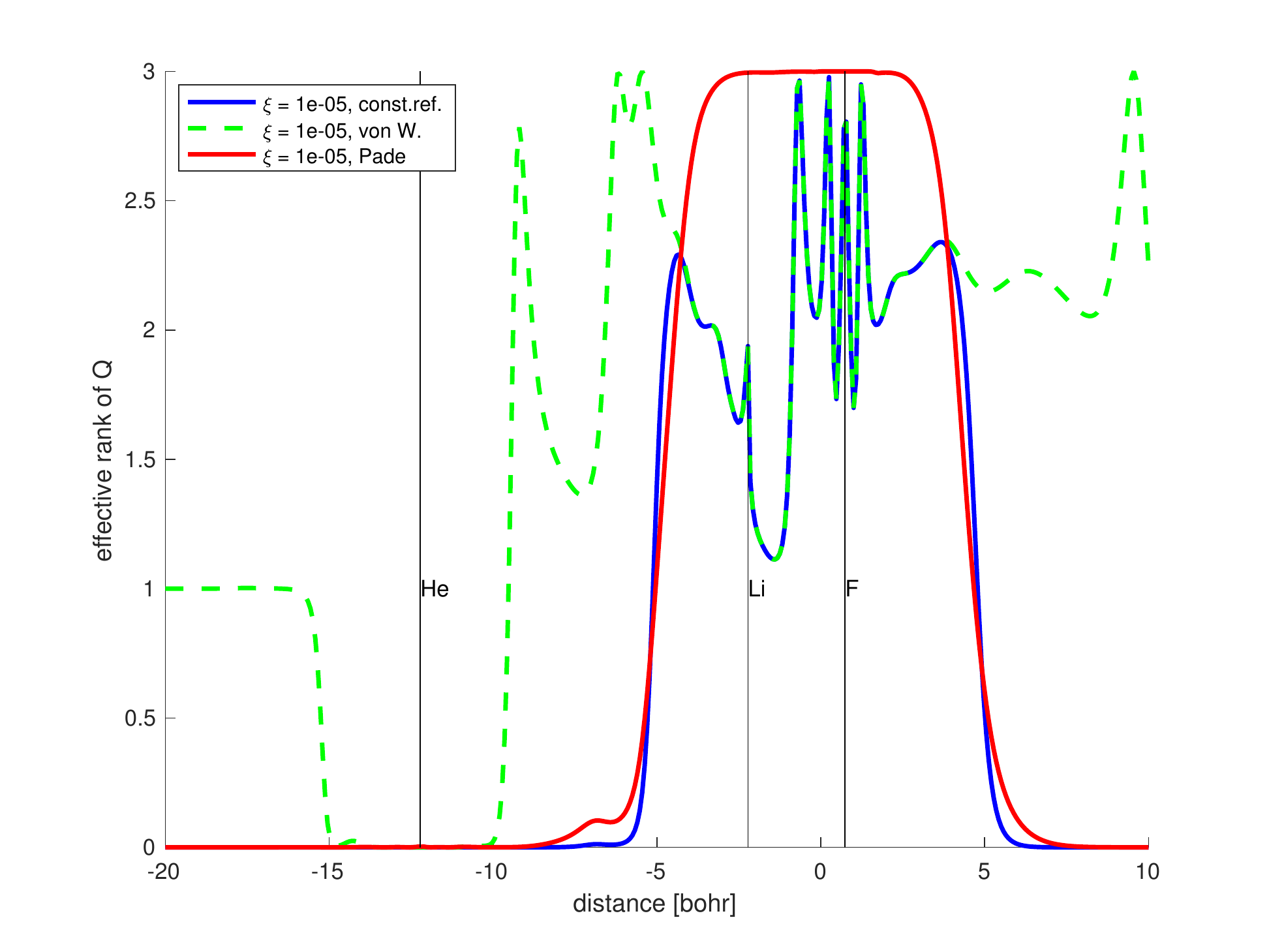}
  \caption{Effective rank estimates at different locations on a line through a helium atom and a LiF molecule. The top and bottom panels show results for $\xi = 10^{-3}$ and $\xi = 10^{-5}$, respectively. The solid blue line shows results for a position-independent reference, the dashed green curve for a von Weizs\"acker reference energy, and the solid red curve shows the Pad\'e approximation.}
  \label{figLineHeLiF}
\end{figure}

\section{The Intrinsic kinetic energy tensor as a generalized isoorbital indicator}

The tensor $\mathbf{Q}$ contains information that goes beyond the kinetic energy density $\tau_{\mathrm{D}}$ and the vorticity alone. Construction of an exchange-correlation functional that exploits this is beyond the scope of the present work. However, in this section we illustrate the additional information by estimating the number $K_{\text{eff}}(\mathbf{r})$ of significant Kohn--Sham orbitals at a given location in space. 

In practical numerical calculations, where small inaccuracies due to numerical noise are always present, the exact matrix rank of $\mathbf{Q}$ is not a useful concept. There are, however, notions of the numerical rank that depend on the singular values of a matrix. For the hermitian, positive semidefinite matrix $\mathbf{Q}$, singular values and eigenvalues $q_l$ coincide. In what follows we assume the order $q_3 \geq q_2 \geq q_1 \geq 0$. Given a threshold $\xi > 0$, a simple numerical rank is the number of singular values that exceed $\xi$. In order to obtain a smooth function of the singular values, we introduce a Pad\'e approximation to the step function,
\begin{equation}
   r_{\mathrm{Pade}}(\mathbf{Q}) = \sum_{l=1}^3 \frac{q_l}{\xi + q_l} = \trace\big( (\mathbf{Q}+\xi I)^{-1} \mathbf{Q} \big).
\end{equation}
Clearly, this numerical rank is lower than the mathematical rank, $r_{\mathrm{Pade}}(\mathbf{Q}) \leq \mathrm{rank}(\mathbf{Q})$. Other numerical rank concepts include the squared ratio of the Frobenius norm to the spectral norm, $\sum_l q_l^2 / q_3^2$, discussed by Rudelson and Vershynin~\cite{RUDELSON_JACM54_21} as well as the effective rank discussed by Roy and Vetterli~\cite{ROY_ESPC_2007}. The latter authors define normalized singular values $p_l = q_l/\trace(\mathbf{Q})$ which are treated as a formal probability distribution to which an entropy measure can be assigned. The original work uses the Shannon entropy $H_1 = -\sum_l p_l \log(p_l)$ and assigns an effective rank $r_1(\mathbf{Q}) = e^{H_1} \geq 1$. In our experience, replacing the Shannon entropy by Renyi entropy of order~2, $H_2 = -\log(p_1^2+p_2^2+p_3^2)$, does not substantially affect the numerical estimates, but yields a particularly simple formula for the effective rank,
\begin{equation}
   r_2(\mathbf{Q}) = e^{H_2} = \frac{\big( \trace(\mathbf{Q}) \big)^2}{\trace(\mathbf{Q}^2)} \geq 1.
\end{equation}
This effective rank can be related to the Hessian of the exchange hole and the vorticity. Recalling Eq.~\eqref{eqQExchHessian} that relates the real, symmetric part $\mathbf{R} = \tfrac{1}{2} (\mathbf{Q} + \mathbf{Q}^*) = \tfrac{1}{2} (\mathbf{Q} + \mathbf{Q}^T)$ to the Hessian of the exchange hole, we note that $\tau_{\mathrm{D}} = \trace(\mathbf{Q}) = \trace(\mathbf{R})$ and $\trace(\mathbf{Q}^2) = \trace(\mathbf{R}^2) + \rho^2 |\nupvec|^2/8$. Hence,
\begin{equation}
   1 \leq r_2(\mathbf{Q}) = \frac{\tau_{\mathrm{D}}^2}{\trace(\mathbf{R}^2) + \rho^2 |\nupvec|^2/8} \leq 3,
\end{equation}
where the inequalities are direct consequences of the properties of the Renyi entropy.
To make the effective rank sensitive to the absolute singular values, rather than just their ratios, we introduce the modified expression
\begin{equation}
   r'_2(\mathbf{Q}) = \frac{\big( \trace(\mathbf{Q}) \big)^2}{\xi^2 \tau_{\text{ref}}^2 + \trace(\mathbf{Q}^2)},
\end{equation}
where $\tau_{\text{ref}}$ is a potentially position-dependent reference energy. Besides a constant $\tau_{\mathrm{ref}} \equiv 1$, we have considered the Thomas--Fermi energy $\tau_{\mathrm{TF}} = C_{\mathrm{TF}} \rho^{5/3}$ and the von Weizs\"acker energy $\tau_{\mathrm{vW}} = \frac{1}{2} |\nabla \sqrt{\rho}|^2$. We find that the results with $\tau_{\mathrm{TF}}$ and $\tau_{\mathrm{vW}}$ as reference are very similar and thus, report only those with $\tau_{\mathrm{vW}}$.

\begin{figure}
  \includegraphics[width=\columnwidth]{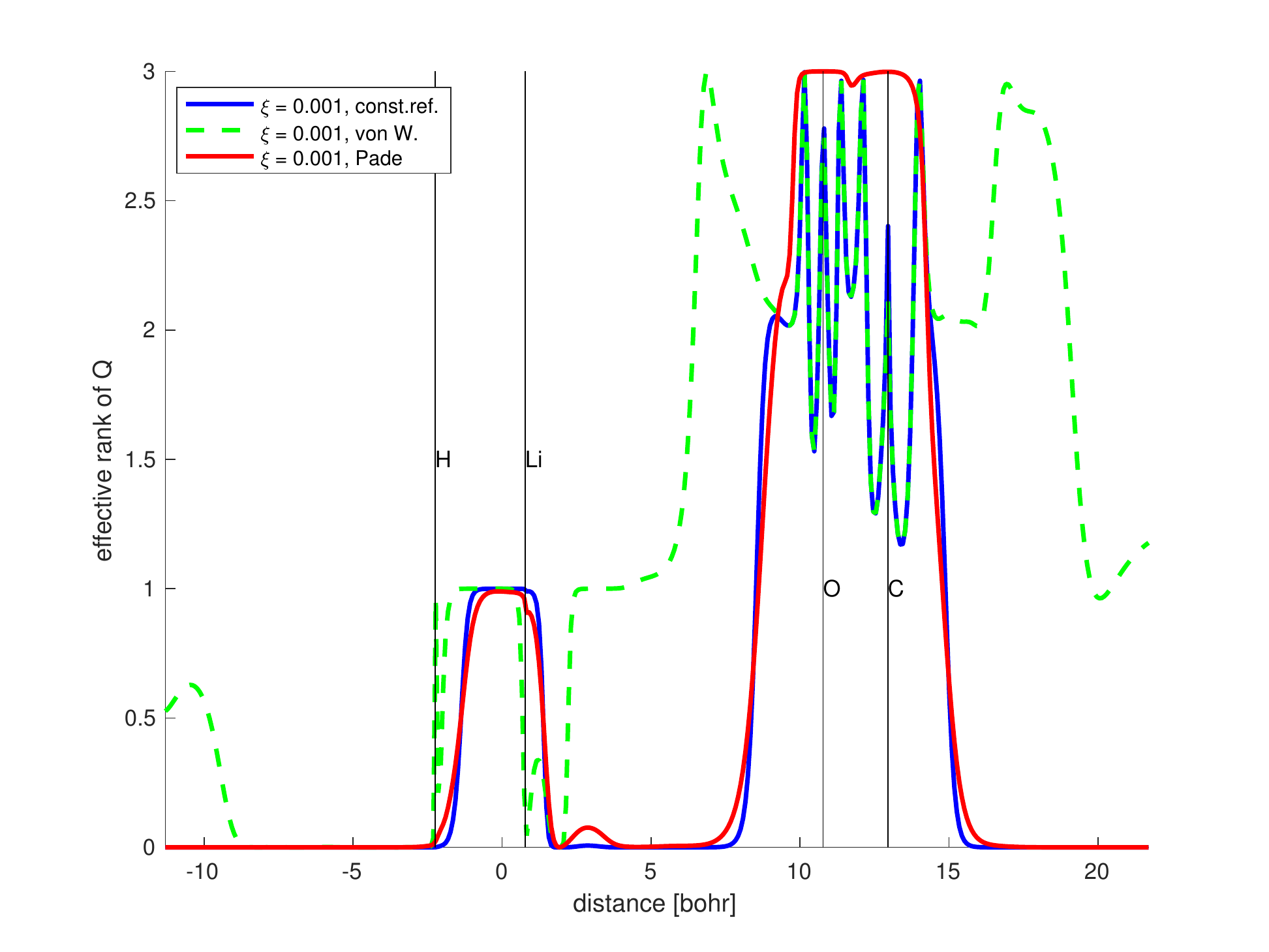} \\
  \includegraphics[width=\columnwidth]{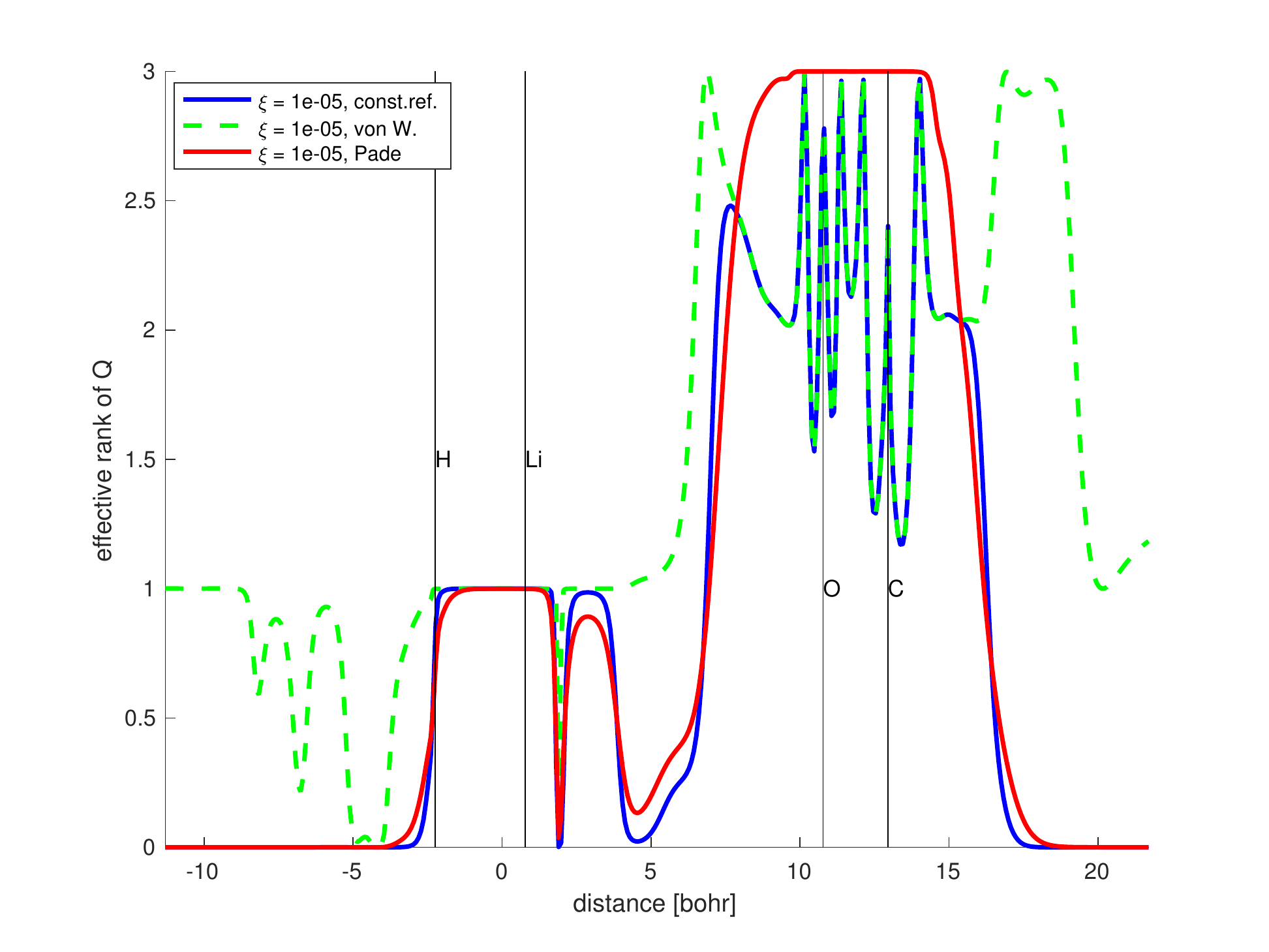}
  \caption{Effective rank estimates at different locations on a line through a the bond axes of a CO and a LiH molecule. The top and bottom panels show results for $\xi = 10^{-3}$ and $\xi = 10^{-5}$, respectively. The solid blue line shows results for a position-independent reference, the dashed green curve for a von Weizs\"acker reference energy, and the solid red curve shows the Pad\'e approximation.}
  \label{figLineCOLiH}
\end{figure}

To investigate the different behavior of the above numerical rank concepts, we obtained the $\mathbf{Q}$ tensor from Kohn--Sham calculations employing the TPSS mGGA functional~\cite{TAO_PRL91_146401} and the aug-cc-pCVTZ basis~\cite{DUNNING_JCP90_1007,WOON_JCP100_2975,KENDALL_JCP96_6796}. London gauge factors were used in calculations at finite magnetic fields~\cite{LONDON_JPR8_397}. All calculations were performed using the DFT implementation~\cite{TELLGREN_JCP140_034101,FURNESS_JCTC11_4169} in the {\sc London} program~\cite{TELLGREN_JCP129_154114,LondonProgram}.

\subsection{Shell structure in the neon atom}

Results for the neon atom are shown in Fig.~\ref{figLineNe}. The high degree of symmetry of this system actually complicates the interpretation in terms of the effective rank of $\mathbf{Q}$. The gradients of the 1s and 2s orbital are parallel as they only have radial components. Hence, together they give only a rank 1 contribution to the canonical tensor $\boldsymbol{\tau}$. A simple 2s orbital of the form $(1-cr) e^{-cr}$ furthermore has a density maximum at at $r=2/c$, where its kinetic energy contribution vanishes. The 2p orbitals generally give a rank 3 contribution, with both radial and angular components, to the canonical tensor $\boldsymbol{\tau}$. However, 2p orbitals of the simple form $xe^{-cr}$, $ye^{-cr}$, and $ze^{-cr}$ have a density maximum at $r=1/c$, where they give a rank 2 contribution because their radial gradients vanish. The von Weizs\"acker-like term, $\boldsymbol{\kappa} \boldsymbol{\kappa}^{\dagger}/(2\rho)$, contributes only a radial gradient, but is not able to completely cancel the orbital contributions. The mathematical rank of $\mathbf{Q}$ is therefore always 3. The effective rank is affected by the varying orders of magnitude of the contributions from the 1s, 2s, and 2p orbitals, which gives rise to oscillations in the effective rank $r'_2(\mathbf{Q})$ that resemble shell structure. A similar phenomenon is well known for the Electron Localization Function, which is also visualized in Fig.~\ref{figLineNe}. The Pad\'e based effective rank, $r_{\mathrm{Pade}}(\mathbf{Q})$ is insensitive to these oscillations and assigns a maximal numerical rank in the whole volume near the neon atom.

\subsection{Examples from intramolecular and asymptotic regions}

The long-range properties are illustrated in Fig.~\ref{figLineNeAr} for a system consisting of a neon atom and an argon atom. Because the 2p orbitals are the slowest decaying orbitals in a neon atom, it possible to argue that the asymptotic region far away from the atom should be considered to be a three-orbital region. However, an alternative perspective is that orbitals should only count when they are non-negligible, so that $\mathbf{Q} \approx \mathbf{0}$ yields an effective rank of approximately zero. The Pad\'e formula $r_{\mathrm{Pade}}(\mathbf{Q})$ is insensitive to the atomic shell structure but gives largely reasonable numerical ranks, which decay to zero far away from any atom. The effective rank $r'_2(\mathbf{Q})$, with a constant $\tau_{\mathrm{ref}} \equiv 1$, has similar asymptotic properties and is in addition sensitive to the shell structure. When the von Weizs\"acker (shown) or Thomas--Fermi reference energies (not shown) are employed, the asymptotic behavior is less controlled in general, although the behavior in Fig.~\ref{figLineNeAr} fits the perspective that the asymptotic region is a three-orbital region (i.e. the effective rank is 2).

Results for a helium atom separated by 10 bohr from a LiF molecule are shown in Fig.~\ref{figLineHeLiF}. The exact locations of the nuclei are $-12.2204$ (He), $-2.2204$ (Li), and $0.7401$~bohr (F). The Pad\'e approximation and the entropy-based $r'_2(\mathbf{Q})$ (with a constant $\tau_{\mathrm{ref}}$) again yield results that assign a vanishing effective rank to the asymptotic region, but differ in the sensitivity to atomic shell structure. The curve for the effective rank $r'_2(\mathbf{Q})$ with von Weizs\"acker reference energy show a plateau to the left of the helium atom, where this measure yields an effective rank of 1 (corresponding to a two-orbital region). This is possible because both the $\mathbf{Q}$ tensor and the von Weizs\"acker (or Thomas--Fermi) energy are negligible in this region.

In Fig.~\ref{figLineCOLiH}, a linear system composed of a LiH and a CO molecule is shown. The coordinates for the nuclei are $-2.2488$ (H), $0.7748$ (Li), $10.7748$ (O), and $12.9486$~bohr (C). The Pad\'e approximation and constant-reference $r'_2(\mathbf{Q})$ again yield vanishing asymptotic rank and high rank in the near the oxygen and carbon atom, but differ regarding shell structure. When the von Weizs\"acker energy is used instead, the asymptotic values are harder to interpret, with oscillations appearing to the left of the LiH molecule when the parameter value is set to $\xi = 10^{-5}$.

\begin{figure}
	\includegraphics[width=\columnwidth]{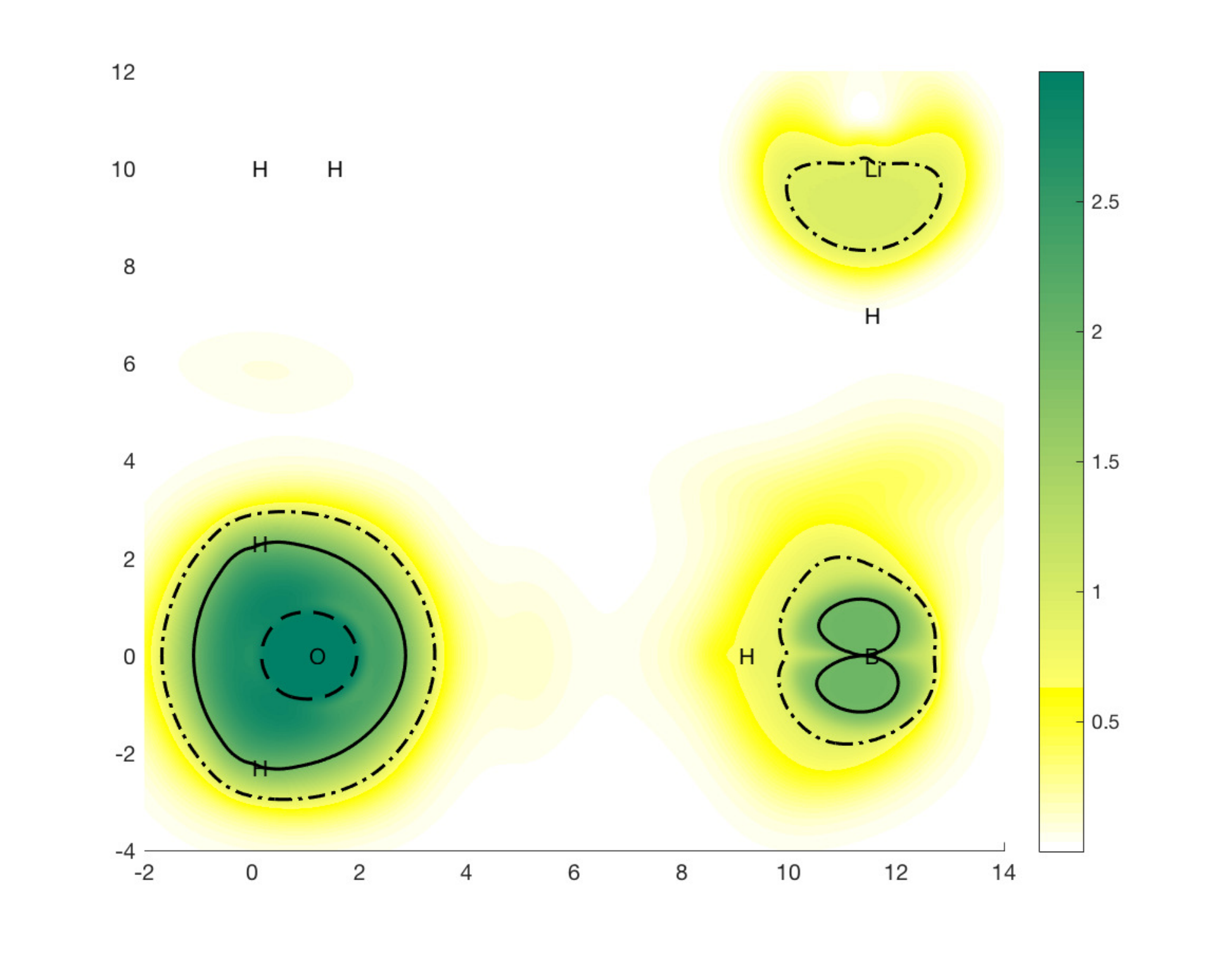} \\
	\includegraphics[width=\columnwidth]{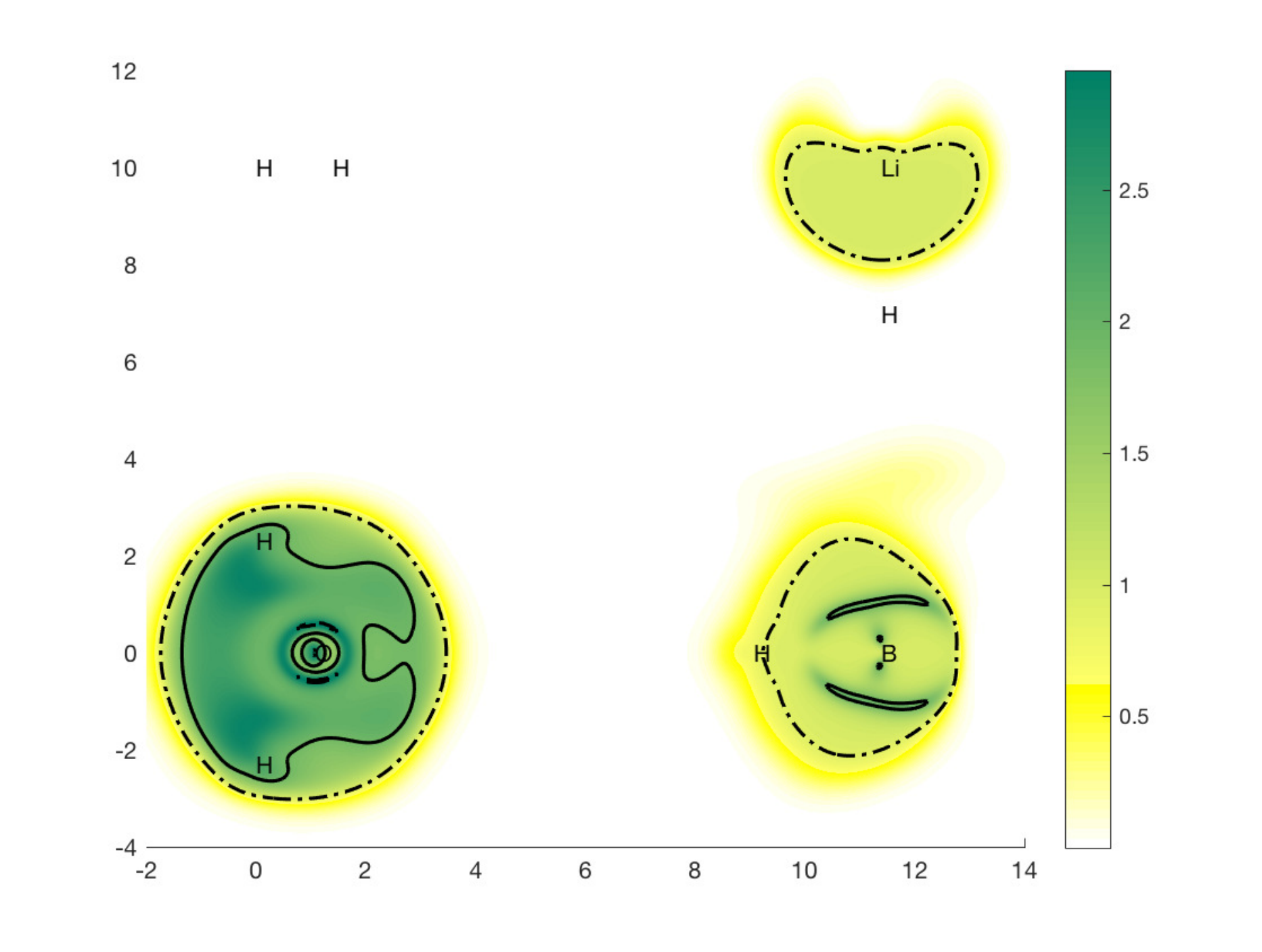}
	\caption{Effective rank estimates for the H$_2$--LiH--BH--H$_2$O system. The top and bottom panels show $r_{\mathrm{Pade}}$ and $r'_2(\mathbf{Q})$, respectively, with the parameter $\xi = 10^{-3}$ and a position-independent reference. Contour lines are displayed for visual support at the levels 0.9 (dash-dot), 1.9 (solid), and 2.9 (dashed).}
	\label{fig4mol_erankB0}
\end{figure}

\subsection{Intra- and intermolecular regions in a four molecule cluster}

A two-dimensional example is shown in Fig.~\ref{fig4mol_erankB0} for a planar supersystem composed of four different small, well-separated molecules: (a) H$_2$ with nuclei placed at $(0,10)$ and $(1.3984,10)$, (b) LiH with nuclei at $(11.3984,10)$ and $(11.3984,6.9764)$, (c) BH with nuclei at $(11.3984,0)$ and $(9.0642,0)$, and (d) H$_2$O with hydrogens at $(0,\pm 2.3010)$ and oxygen at $(1.0690,0)$. All coordinates are in units of bohr. These subsystems contribute a one-, two-, three-, and five-orbital region, respectively. The parameter value $\xi=10^{-3}$~au and a constant $\tau_{\mathrm{ref}} \equiv 1$ was used to produce the plot. As expected, the H$_2$ molecule, being a one-orbital system in isolation, is invisible in effective rank plots. Likewise, for both $r_{\mathrm{Pade}}$ and $r'_2$, the peak value in the region inside the LiH molecule is 1, as expected for a two-orbital region. In the BH molecule, even the Pade-based numerical rank shows non-trivial structure with two peaks near 2---indicating three-orbital regions---on either side of the bond axis. In the plot of $r'_2$, the peak regions with values near 2 are much more narrow and elongated. Finally, in the region around the H$_2$O molecule, the Pad\'e-based rank shows a peak of maximum rank near the oxygen atom, and regular decay. The indicated four-, three-, and two-orbital regions roughly have the shapes of concentric circular disks. When the numerical rank is based on $r'_2$, the two-orbital region in similar, but the three- and four-orbital regions are more complicated due to shell structure and orbital oscillations inside the molecule.

\begin{figure}
  \includegraphics[width=\columnwidth]{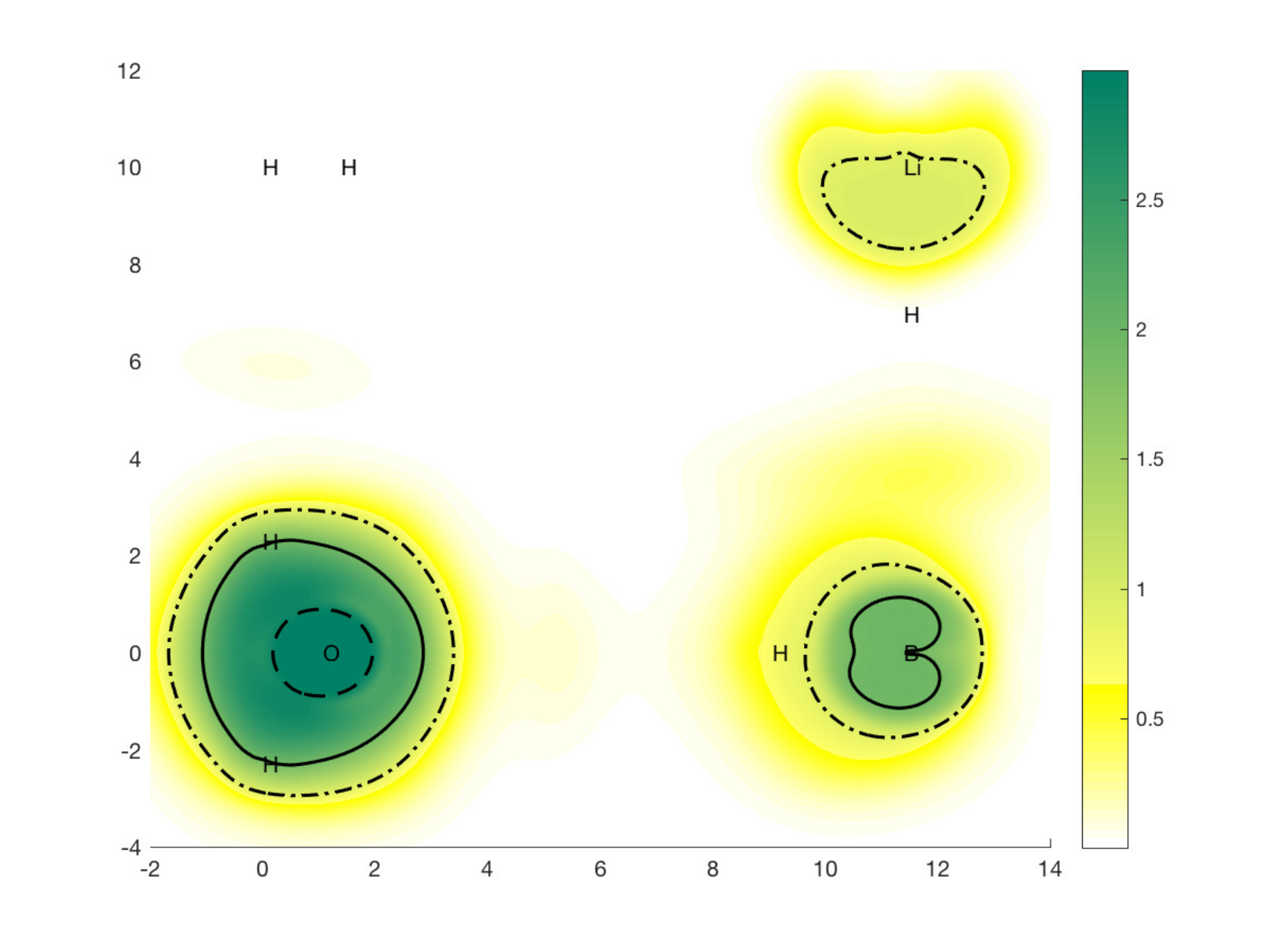} \\
	\includegraphics[width=\columnwidth]{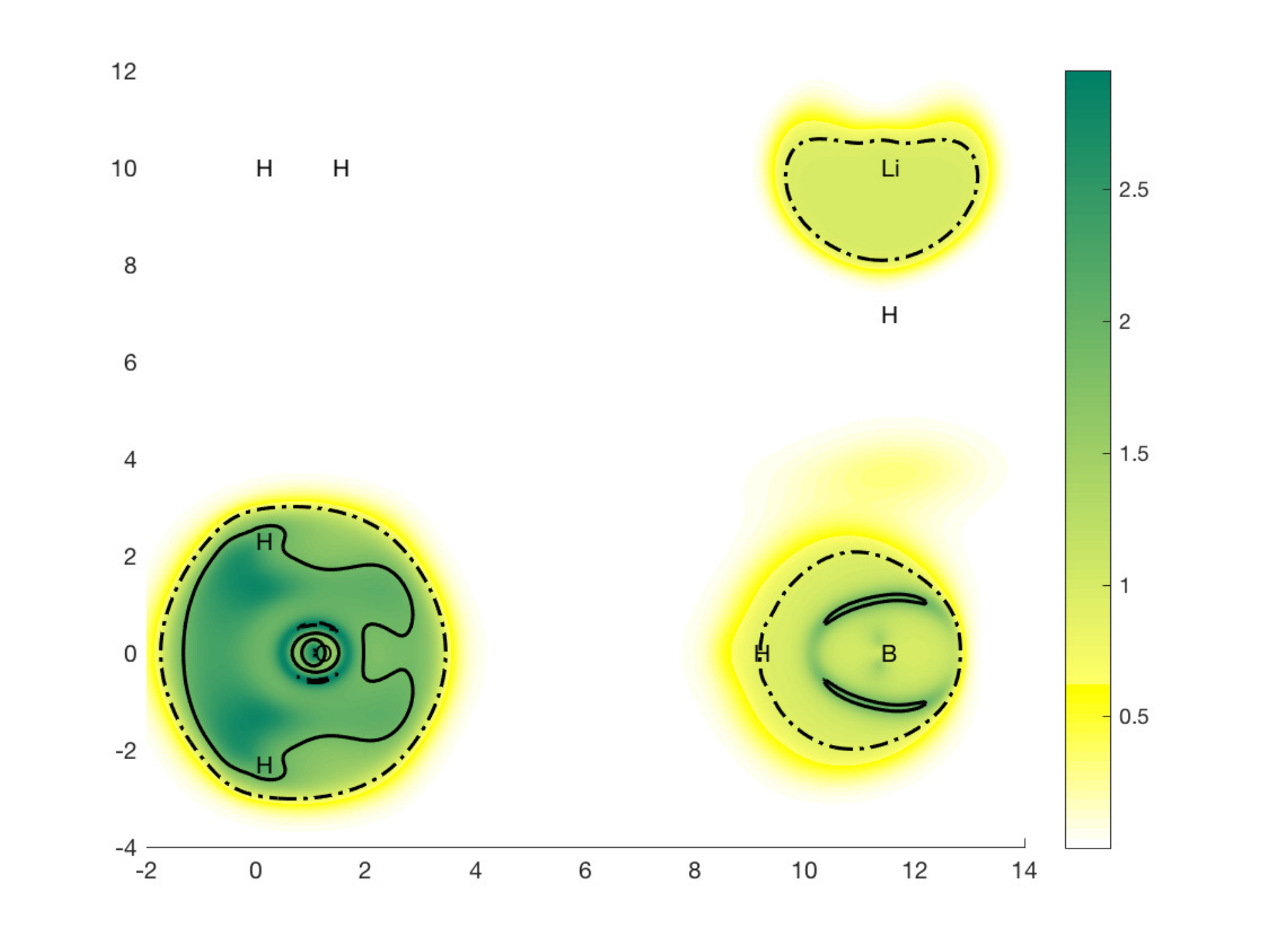}
	\caption{Effective rank estimates for the H$_2$--LiH--BH--H$_2$O system in a magnetic field of 0.1~a.u. perpendicular to the molecular plane. The top and bottom panels show results for the $r_{\mathrm{Pade}}$ and $r'_2(\mathbf{Q})$, respectively, with $\xi = 10^{-3}$ and a position-independent reference. Contour lines are displayed for visual support at the levels 0.9 (dash-dot), 1.9 (solid), and 2.9 (dashed).}
	\label{fig4mol_erankB0.1z}
\end{figure}

\begin{figure}
  \includegraphics[width=\columnwidth]{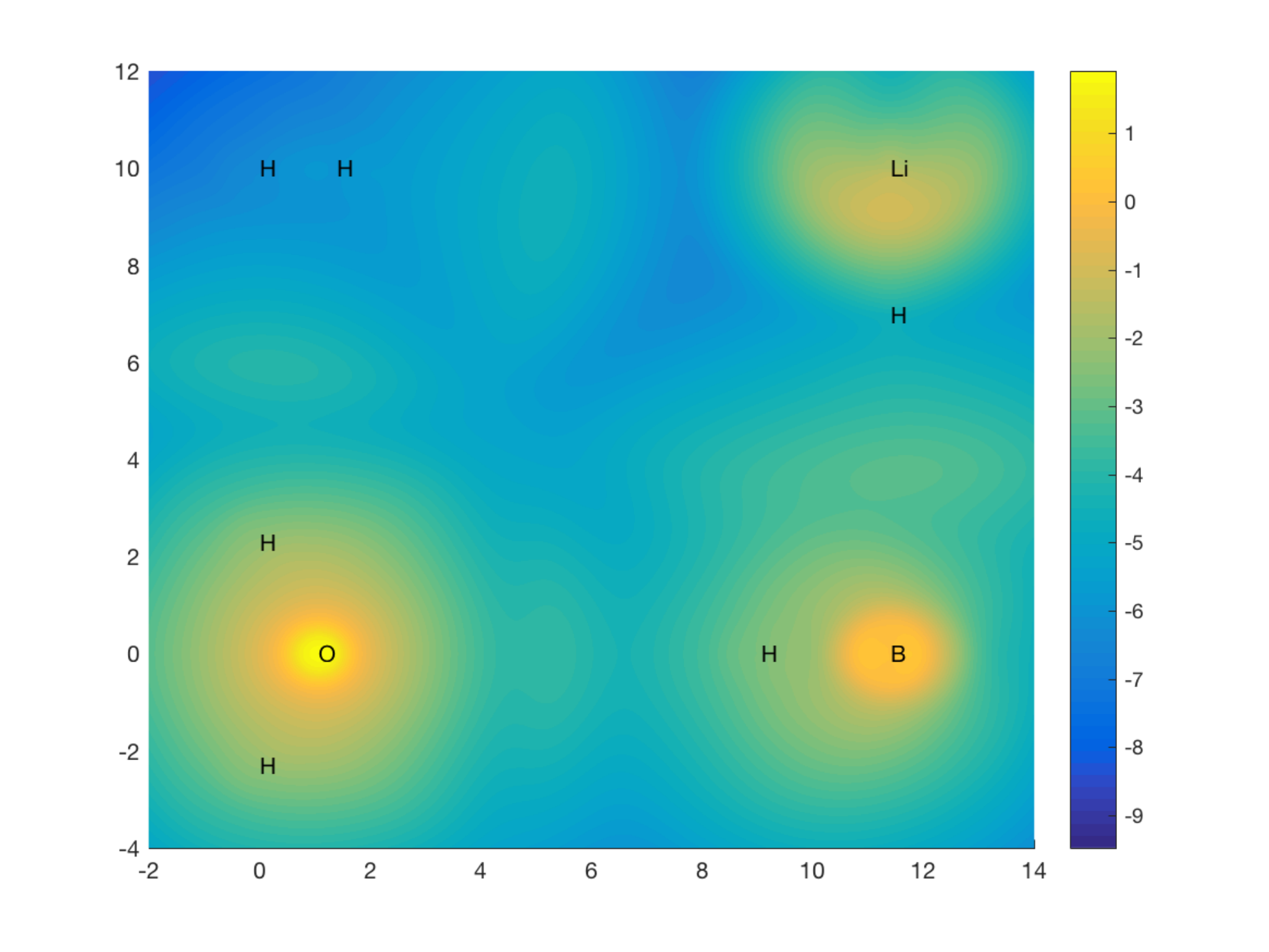}  \\
  \includegraphics[width=\columnwidth]{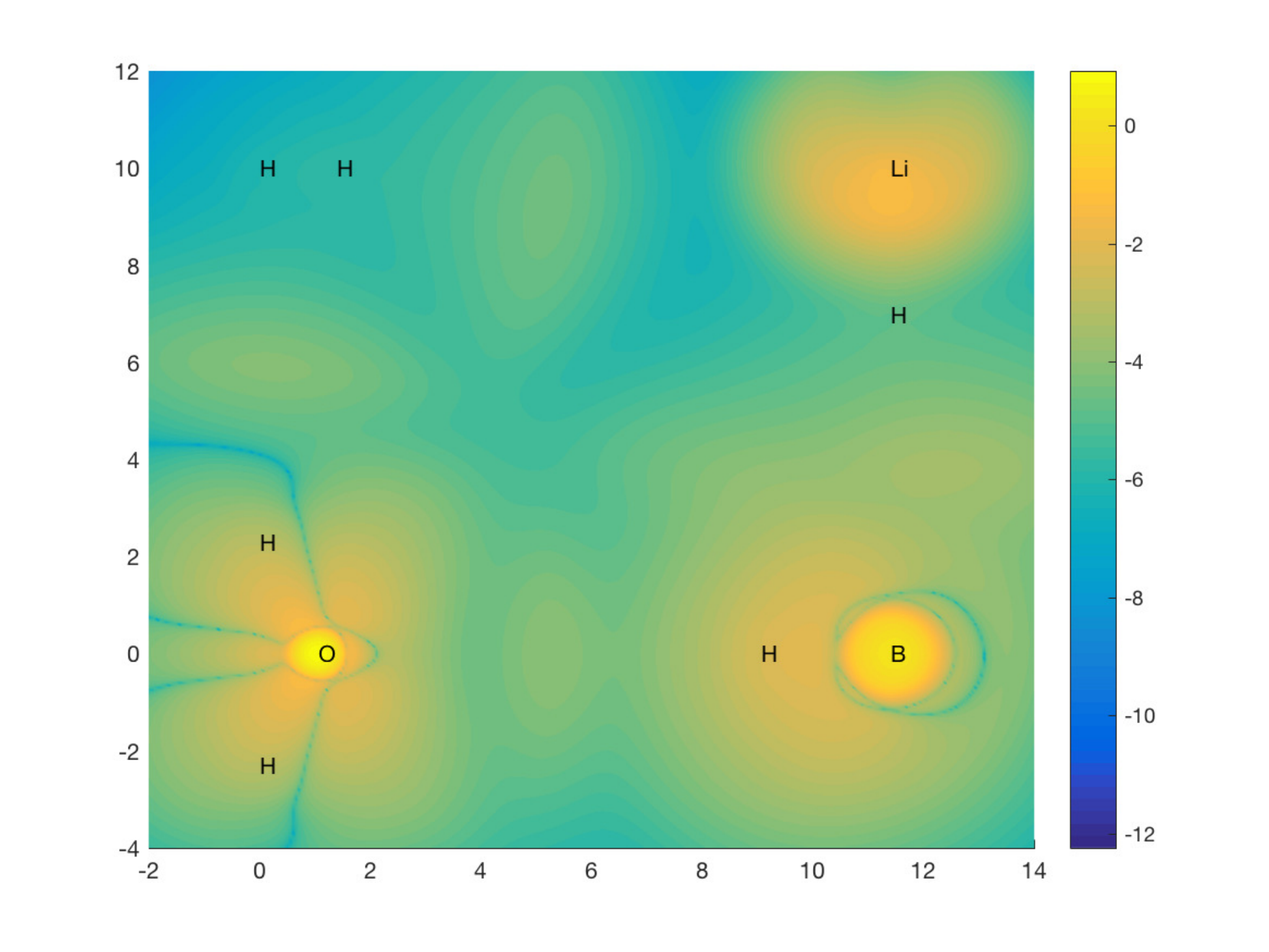} 
	\caption{ The top and bottom panels show $\log_{10}(\tau_D)$ and $\log_{10}(\rho|\boldsymbol{\nu}|)$, respectively, for the H$_2$--LiH--BH--H$_2$O system in a perpendicular magnetic field of 0.1~a.u. Note that $\tau_{\mathrm{D}}$ and $\rho|\nupvec|$ are qualitatively very similar, but differ in the finer details, such as the exact height and location of the peak values.}
	\label{fig4mol_tauDVortB0.1z}
\end{figure}

\begin{figure}
	\includegraphics[width=\columnwidth]{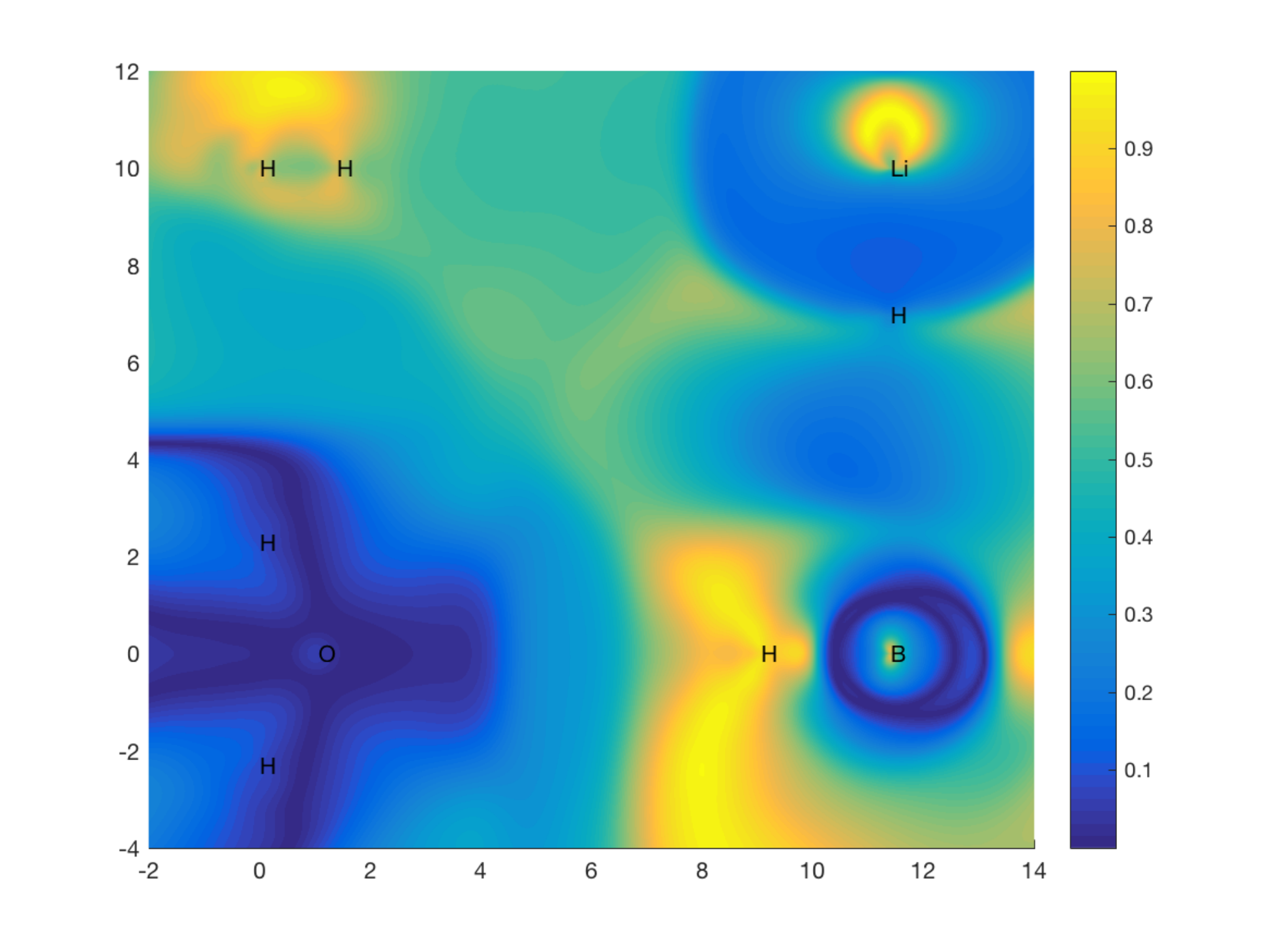}
	\caption{ The ratio $\rho|\boldsymbol{\nu}|/2\tau_D$ is plotted for the H$_2$--LiH--BH--H$_2$O system in a perpendicular magnetic field of 0.1~a.u. Note that the bound in Eq.~\eqref{eqNuTauDBound} is satisfied everywhere and that both $\tau_{\mathrm{D}}$ and $\rho|\nupvec|$ are in tiny in some of the regions of near maximal ratio.}
	\label{fig4mol_ratio_tauDVortB0.1z}
\end{figure}

The same supersystem of four molecules has been subjected to a perpendicular magnetic field $B_{\perp} = 0.1$~au in Fig.~\ref{fig4mol_erankB0.1z}. Although this is a very strong field compared what is accessible experimentally, the effects on the electronic structure and the density are fairly subtle. However, the singlet state in the BH molecule is known to change character from diamagnetic to paramagnetic around 0.2~au~\cite{TELLGREN_JCP129_154114,TELLGREN_PCCP11_5489}. In the Pad\'e-based plot, the main visible effect is that the two separate peaks of near-maximal rank 2 have merged to a single region and there is no longer a pronounced valley on the bond axis. In the $r'_2$ plot, the contour line at 0.9 has a more regular shape compared to the zero-field case.

\subsection{Illustration of the vorticity bound}

The perpendicular magnetic field induces a vorticity, which is visualized alongside Dobson's kinetic energy density in Fig.~\ref{fig4mol_ratio_tauDVortB0.1z}. Distributions of $\tau_{\mathrm{D}}$ and $\rho|\nupvec|$ are qualitatively very similar, though their detailed structure differ. In particular, they have their peaks at slightly different positions. In Fig.~\ref{fig4mol_ratio_tauDVortB0.1z}, we plot the ratio
\begin{equation*}
   \frac{\rho |\nupvec|}{2 \tau_{\mathrm{D}}}.
\end{equation*}
The fact that this ratio nowhere exceeds 1 provides a numerical illustration of the universal bound in Eq.~\eqref{eqNuTauDBound} above. Note that both the numerator and denominator are very small in some regions of near maximal ratio.

\section{Discussion and conclusion}

We have presented a kinetic energy tensor which contains Dobson's kinetic energy density and the paramagnetic vorticity as essentially independent components. More precisely, Dobson's $\tau_{\mathrm{D}}$ is the trace (sum of diagonal elements), whereas the vector $\rho\nupvec$ is encoded in the imaginary, anti-symmetric part. From fact that $\mathbf{Q}$ is positive semidefinite we discover a new bound that the vorticity cannot exceed: $\rho|\nupvec| \leq 2 \tau_{\mathrm{D}}$. In light of this, it is natural to place vorticity-dependent exchange-correlations alongside mGGAs on the third rung of Jacob's ladder. Furthermore, it is natural to expand the third rung to a $\mathbf{Q}$-dependent form that subsumes both mGGAs and vorticity-dependent functionals as special cases. The upper bound on the vorticity also raises the possibility of defining a vorticity-corrected scalar kinetic energy density. Whereas Dobson's $\tau_{\mathrm{D}}$ is mainly a gauge-corrected kinetic energy density, the scalar density
\begin{equation}
  \tau_{\mathrm{V}} = \tau_{\mathrm{can}} - \frac{|\nabla\rho|^2}{8\rho} - \frac{|\jpvec|^2}{2\rho} - \frac{\rho |\nupvec|}{2} \geq 0
\end{equation}
builds in a vorticity dependence while retaining non-negativity universally for all systems. Another conceivable approach to incorporate a vorticity-dependence into mGGAs is to replace the trace norm $\tau_{\mathrm{D}} = \|\mathbf{Q}\|_1$ by some other norm in the mGGA form (Eq.~\eqref{eqMGGAForm}).

The very clearcut $N$-representability conditions on the intrinsic kinetic energy tensor $\mathbf{Q}$ furthermore relate its matrix rank to the number of significant orbitals. By defining a numerical effective rank, which roughly stated filters out small eigenvalues of $\mathbf{Q}$, it is possible to define a position-dependent effective rank that yields a count of the number of non-negligible orbitals at given locations. Our numerical results above demonstrate that such an effective orbital count can distinguish between, for example, the interior of a LiH molecule and a BH molecule based on local information. There is a wide range of possibilities for tuning effective ranks and the related orbital counts to visualize chemical information. We have exemplified this by computing a Pad\'e-based rank, which is insensitive to atomic shell structure, and an entropy-based rank, which is sensitive to shell structure. These effective orbital counts are broadly related to quantities such as the Electron Localization Function~\cite{BECKE_JCP92_5397,SILVI_N371_683} and its current-corrected form~\cite{FURNESS_MP114_1415}, quantum stress~\cite{TAO_PRL100_206405,TAO_PRB92_060401}, and various uses of stress tensors for chemical interpretation~\cite{BADER_JCP73_2871,GUEVARAGARCIA_JCP134_234106}. However, the information available in the $\mathbf{Q}$ tensor goes beyond scalar densities and has a different character than other tensor densities due to the crisp $N$-representability conditions and the relation to exchange hole curvature tensor. As recent work has seen renewed efforts to develop functionals for two- and few-electron systems~\cite{ENTWISTLE_PRB94_205134,SUN_JCP144_19101, SUN_JCP145_019902}, the present results raise the prospect of using the information in $\mathbf{Q}$ to construct a local interpolation of different functionals adapted for one-, two-, three-, and many-orbital systems, respectively.

\section*{Acknowledgments}

This work was supported by the Research Council of Norway through Grant No.~240674 and CoE Hylleraas Centre for Molecular Sciences Grant No.~262695, and the European  Union’s Horizon 2020 research and innovation programme under the Marie Sk{\l}odowska-Curie grant agreement No.~745336. This work has also received support from the Norwegian Supercomputing Program (NOTUR) through a grant of computer time (Grant No.~NN4654K). We thank A.~M.~Teale for useful discussions.


\begin{thebibliography}{47}%
\makeatletter
\providecommand \@ifxundefined [1]{%
 \@ifx{#1\undefined}
}%
\providecommand \@ifnum [1]{%
 \ifnum #1\expandafter \@firstoftwo
 \else \expandafter \@secondoftwo
 \fi
}%
\providecommand \@ifx [1]{%
 \ifx #1\expandafter \@firstoftwo
 \else \expandafter \@secondoftwo
 \fi
}%
\providecommand \natexlab [1]{#1}%
\providecommand \enquote  [1]{``#1''}%
\providecommand \bibnamefont  [1]{#1}%
\providecommand \bibfnamefont [1]{#1}%
\providecommand \citenamefont [1]{#1}%
\providecommand \href@noop [0]{\@secondoftwo}%
\providecommand \href [0]{\begingroup \@sanitize@url \@href}%
\providecommand \@href[1]{\@@startlink{#1}\@@href}%
\providecommand \@@href[1]{\endgroup#1\@@endlink}%
\providecommand \@sanitize@url [0]{\catcode `\\12\catcode `\$12\catcode
  `\&12\catcode `\#12\catcode `\^12\catcode `\_12\catcode `\%12\relax}%
\providecommand \@@startlink[1]{}%
\providecommand \@@endlink[0]{}%
\providecommand \url  [0]{\begingroup\@sanitize@url \@url }%
\providecommand \@url [1]{\endgroup\@href {#1}{\urlprefix }}%
\providecommand \urlprefix  [0]{URL }%
\providecommand \Eprint [0]{\href }%
\providecommand \doibase [0]{http://dx.doi.org/}%
\providecommand \selectlanguage [0]{\@gobble}%
\providecommand \bibinfo  [0]{\@secondoftwo}%
\providecommand \bibfield  [0]{\@secondoftwo}%
\providecommand \translation [1]{[#1]}%
\providecommand \BibitemOpen [0]{}%
\providecommand \bibitemStop [0]{}%
\providecommand \bibitemNoStop [0]{.\EOS\space}%
\providecommand \EOS [0]{\spacefactor3000\relax}%
\providecommand \BibitemShut  [1]{\csname bibitem#1\endcsname}%
\let\auto@bib@innerbib\@empty
%</preamble>
\bibitem [{\citenamefont {Vignale}\ and\ \citenamefont
  {Rasolt}(1987)}]{VIGNALE_PRL59_2360}%
  \BibitemOpen
  \bibfield  {author} {\bibinfo {author} {\bibfnamefont {G.}~\bibnamefont
  {Vignale}}\ and\ \bibinfo {author} {\bibfnamefont {M.}~\bibnamefont
  {Rasolt}},\ }\href {\doibase 10.1103/PhysRevLett.59.2360} {\bibfield
  {journal} {\bibinfo  {journal} {Phys. Rev. Lett.}\ }\textbf {\bibinfo
  {volume} {59}},\ \bibinfo {pages} {2360} (\bibinfo {year}
  {1987})}\BibitemShut {NoStop}%
\bibitem [{\citenamefont {Capelle}\ and\ \citenamefont
  {Vignale}(2002)}]{CAPELLE_PRB65_113106}%
  \BibitemOpen
  \bibfield  {author} {\bibinfo {author} {\bibfnamefont {K.}~\bibnamefont
  {Capelle}}\ and\ \bibinfo {author} {\bibfnamefont {G.}~\bibnamefont
  {Vignale}},\ }\href {\doibase 10.1103/PhysRevB.65.113106} {\bibfield
  {journal} {\bibinfo  {journal} {Phys. Rev. B}\ }\textbf {\bibinfo {volume}
  {65}},\ \bibinfo {pages} {113106} (\bibinfo {year} {2002})}\BibitemShut
  {NoStop}%
\bibitem [{\citenamefont {Vignale}\ \emph {et~al.}(1988)\citenamefont
  {Vignale}, \citenamefont {Rasolt},\ and\ \citenamefont
  {Geldart}}]{VIGNALE_PRB37_2502}%
  \BibitemOpen
  \bibfield  {author} {\bibinfo {author} {\bibfnamefont {G.}~\bibnamefont
  {Vignale}}, \bibinfo {author} {\bibfnamefont {M.}~\bibnamefont {Rasolt}}, \
  and\ \bibinfo {author} {\bibfnamefont {D.~J.~W.}\ \bibnamefont {Geldart}},\
  }\href {\doibase 10.1103/PhysRevB.37.2502} {\bibfield  {journal} {\bibinfo
  {journal} {Phys. Rev. B}\ }\textbf {\bibinfo {volume} {37}},\ \bibinfo
  {pages} {2502} (\bibinfo {year} {1988})}\BibitemShut {NoStop}%
\bibitem [{\citenamefont {Skudlarski}\ and\ \citenamefont
  {Vignale}(1993)}]{SKUDLARSKI_PRB48_8547}%
  \BibitemOpen
  \bibfield  {author} {\bibinfo {author} {\bibfnamefont {P.}~\bibnamefont
  {Skudlarski}}\ and\ \bibinfo {author} {\bibfnamefont {G.}~\bibnamefont
  {Vignale}},\ }\href {\doibase 10.1103/PhysRevB.48.8547} {\bibfield  {journal}
  {\bibinfo  {journal} {Phys. Rev. B}\ }\textbf {\bibinfo {volume} {48}},\
  \bibinfo {pages} {8547} (\bibinfo {year} {1993})}\BibitemShut {NoStop}%
\bibitem [{\citenamefont {Lee}\ \emph {et~al.}(1995)\citenamefont {Lee},
  \citenamefont {Handy},\ and\ \citenamefont {Colwell}}]{LEE_JCP103_10095}%
  \BibitemOpen
  \bibfield  {author} {\bibinfo {author} {\bibfnamefont {A.~M.}\ \bibnamefont
  {Lee}}, \bibinfo {author} {\bibfnamefont {N.~C.}\ \bibnamefont {Handy}}, \
  and\ \bibinfo {author} {\bibfnamefont {S.~M.}\ \bibnamefont {Colwell}},\
  }\href@noop {} {\bibfield  {journal} {\bibinfo  {journal} {J. Chem. Phys.}\
  }\textbf {\bibinfo {volume} {103}},\ \bibinfo {pages} {10095} (\bibinfo
  {year} {1995})}\BibitemShut {NoStop}%
\bibitem [{\citenamefont {Tao}\ and\ \citenamefont
  {Vignale}(2006)}]{TAO_PRB74_193108}%
  \BibitemOpen
  \bibfield  {author} {\bibinfo {author} {\bibfnamefont {J.}~\bibnamefont
  {Tao}}\ and\ \bibinfo {author} {\bibfnamefont {G.}~\bibnamefont {Vignale}},\
  }\href {\doibase 10.1103/PhysRevB.74.193108} {\bibfield  {journal} {\bibinfo
  {journal} {Phys. Rev. B}\ }\textbf {\bibinfo {volume} {74}},\ \bibinfo
  {pages} {193108} (\bibinfo {year} {2006})}\BibitemShut {NoStop}%
\bibitem [{\citenamefont {Tellgren}\ \emph
  {et~al.}(2014{\natexlab{a}})\citenamefont {Tellgren}, \citenamefont {Teale},
  \citenamefont {Furness}, \citenamefont {Lange}, \citenamefont {Ekstr{\"o}m},\
  and\ \citenamefont {Helgaker}}]{TELLGREN_JCP140_034101}%
  \BibitemOpen
  \bibfield  {author} {\bibinfo {author} {\bibfnamefont {E.~I.}\ \bibnamefont
  {Tellgren}}, \bibinfo {author} {\bibfnamefont {A.~M.}\ \bibnamefont {Teale}},
  \bibinfo {author} {\bibfnamefont {J.~W.}\ \bibnamefont {Furness}}, \bibinfo
  {author} {\bibfnamefont {K.~K.}\ \bibnamefont {Lange}}, \bibinfo {author}
  {\bibfnamefont {U.}~\bibnamefont {Ekstr{\"o}m}}, \ and\ \bibinfo {author}
  {\bibfnamefont {T.}~\bibnamefont {Helgaker}},\ }\href@noop {} {\bibfield
  {journal} {\bibinfo  {journal} {J. Chem. Phys.}\ }\textbf {\bibinfo {volume}
  {140}},\ \bibinfo {eid} {034101} (\bibinfo {year}
  {2014}{\natexlab{a}})}\BibitemShut {NoStop}%
\bibitem [{\citenamefont {Perdew}\ and\ \citenamefont
  {Schmidt}(2001)}]{PERDEW_AIPCP577_1}%
  \BibitemOpen
  \bibfield  {author} {\bibinfo {author} {\bibfnamefont {J.~P.}\ \bibnamefont
  {Perdew}}\ and\ \bibinfo {author} {\bibfnamefont {K.}~\bibnamefont
  {Schmidt}},\ }\href {\doibase 10.1063/1.1390175} {\bibfield  {journal}
  {\bibinfo  {journal} {AIP Conf. Proc.}\ }\textbf {\bibinfo {volume} {577}},\
  \bibinfo {pages} {1} (\bibinfo {year} {2001})}\BibitemShut {NoStop}%
\bibitem [{\citenamefont {Becke}\ and\ \citenamefont
  {Roussel}(1989)}]{BECKE_PRA39_3761}%
  \BibitemOpen
  \bibfield  {author} {\bibinfo {author} {\bibfnamefont {A.~D.}\ \bibnamefont
  {Becke}}\ and\ \bibinfo {author} {\bibfnamefont {M.~R.}\ \bibnamefont
  {Roussel}},\ }\href {\doibase 10.1103/PhysRevA.39.3761} {\bibfield  {journal}
  {\bibinfo  {journal} {Phys. Rev. A}\ }\textbf {\bibinfo {volume} {39}},\
  \bibinfo {pages} {3761} (\bibinfo {year} {1989})}\BibitemShut {NoStop}%
\bibitem [{\citenamefont {Perdew}\ \emph {et~al.}(1999)\citenamefont {Perdew},
  \citenamefont {Kurth}, \citenamefont {Zupan},\ and\ \citenamefont
  {Blaha}}]{PERDEW_PRL82_2544}%
  \BibitemOpen
  \bibfield  {author} {\bibinfo {author} {\bibfnamefont {J.~P.}\ \bibnamefont
  {Perdew}}, \bibinfo {author} {\bibfnamefont {S.}~\bibnamefont {Kurth}},
  \bibinfo {author} {\bibfnamefont {A.}~\bibnamefont {Zupan}}, \ and\ \bibinfo
  {author} {\bibfnamefont {P.}~\bibnamefont {Blaha}},\ }\href {\doibase
  10.1103/PhysRevLett.82.2544} {\bibfield  {journal} {\bibinfo  {journal}
  {Phys. Rev. Lett.}\ }\textbf {\bibinfo {volume} {82}},\ \bibinfo {pages}
  {2544} (\bibinfo {year} {1999})}\BibitemShut {NoStop}%
\bibitem [{\citenamefont {Jansen}(1991)}]{JANSEN_PRB43_12025}%
  \BibitemOpen
  \bibfield  {author} {\bibinfo {author} {\bibfnamefont {H.~J.~F.}\
  \bibnamefont {Jansen}},\ }\href {\doibase 10.1103/PhysRevB.43.12025}
  {\bibfield  {journal} {\bibinfo  {journal} {Phys. Rev. B}\ }\textbf {\bibinfo
  {volume} {43}},\ \bibinfo {pages} {12025} (\bibinfo {year}
  {1991})}\BibitemShut {NoStop}%
\bibitem [{\citenamefont {Higuchi}\ and\ \citenamefont
  {Higuchi}(2004)}]{HIGUCHI_PRB69_035113}%
  \BibitemOpen
  \bibfield  {author} {\bibinfo {author} {\bibfnamefont {M.}~\bibnamefont
  {Higuchi}}\ and\ \bibinfo {author} {\bibfnamefont {K.}~\bibnamefont
  {Higuchi}},\ }\href {\doibase 10.1103/PhysRevB.69.035113} {\bibfield
  {journal} {\bibinfo  {journal} {Phys. Rev. B}\ }\textbf {\bibinfo {volume}
  {69}},\ \bibinfo {pages} {035113} (\bibinfo {year} {2004})}\BibitemShut
  {NoStop}%
\bibitem [{\citenamefont {Ayers}\ and\ \citenamefont
  {Fuentealba}(2009)}]{AYERS_PRA80_032510}%
  \BibitemOpen
  \bibfield  {author} {\bibinfo {author} {\bibfnamefont {P.~W.}\ \bibnamefont
  {Ayers}}\ and\ \bibinfo {author} {\bibfnamefont {P.}~\bibnamefont
  {Fuentealba}},\ }\href {http://link.aps.org/abstract/PRA/v80/e032510}
  {\bibfield  {journal} {\bibinfo  {journal} {Phys. Rev. A}\ }\textbf {\bibinfo
  {volume} {80}},\ \bibinfo {eid} {032510} (\bibinfo {year}
  {2009})}\BibitemShut {NoStop}%
\bibitem [{\citenamefont {Ayers}\ and\ \citenamefont
  {Nagy}(2007)}]{AYERS_JCP126_144108}%
  \BibitemOpen
  \bibfield  {author} {\bibinfo {author} {\bibfnamefont {P.~W.}\ \bibnamefont
  {Ayers}}\ and\ \bibinfo {author} {\bibfnamefont {A.}~\bibnamefont {Nagy}},\
  }\href@noop {} {\bibfield  {journal} {\bibinfo  {journal} {J. Chem. Phys.}\
  }\textbf {\bibinfo {volume} {126}},\ \bibinfo {pages} {144108} (\bibinfo
  {year} {2007})}\BibitemShut {NoStop}%
\bibitem [{\citenamefont {Tao}\ \emph {et~al.}(2003)\citenamefont {Tao},
  \citenamefont {Perdew}, \citenamefont {Staroverov},\ and\ \citenamefont
  {Scuseria}}]{TAO_PRL91_146401}%
  \BibitemOpen
  \bibfield  {author} {\bibinfo {author} {\bibfnamefont {J.}~\bibnamefont
  {Tao}}, \bibinfo {author} {\bibfnamefont {J.~P.}\ \bibnamefont {Perdew}},
  \bibinfo {author} {\bibfnamefont {V.~N.}\ \bibnamefont {Staroverov}}, \ and\
  \bibinfo {author} {\bibfnamefont {G.~E.}\ \bibnamefont {Scuseria}},\ }\href
  {\doibase 10.1103/PhysRevLett.91.146401} {\bibfield  {journal} {\bibinfo
  {journal} {Phys. Rev. Lett.}\ }\textbf {\bibinfo {volume} {91}},\ \bibinfo
  {pages} {146401} (\bibinfo {year} {2003})}\BibitemShut {NoStop}%
\bibitem [{\citenamefont {Sun}\ \emph {et~al.}(2013)\citenamefont {Sun},
  \citenamefont {Xiao}, \citenamefont {Fang}, \citenamefont {Haunschild},
  \citenamefont {Hao}, \citenamefont {Ruzsinszky}, \citenamefont {Csonka},
  \citenamefont {Scuseria},\ and\ \citenamefont {Perdew}}]{SUN_PRL111_106401}%
  \BibitemOpen
  \bibfield  {author} {\bibinfo {author} {\bibfnamefont {J.}~\bibnamefont
  {Sun}}, \bibinfo {author} {\bibfnamefont {B.}~\bibnamefont {Xiao}}, \bibinfo
  {author} {\bibfnamefont {Y.}~\bibnamefont {Fang}}, \bibinfo {author}
  {\bibfnamefont {R.}~\bibnamefont {Haunschild}}, \bibinfo {author}
  {\bibfnamefont {P.}~\bibnamefont {Hao}}, \bibinfo {author} {\bibfnamefont
  {A.}~\bibnamefont {Ruzsinszky}}, \bibinfo {author} {\bibfnamefont {G.~I.}\
  \bibnamefont {Csonka}}, \bibinfo {author} {\bibfnamefont {G.~E.}\
  \bibnamefont {Scuseria}}, \ and\ \bibinfo {author} {\bibfnamefont {J.~P.}\
  \bibnamefont {Perdew}},\ }\href {\doibase 10.1103/PhysRevLett.111.106401}
  {\bibfield  {journal} {\bibinfo  {journal} {Phys. Rev. Lett.}\ }\textbf
  {\bibinfo {volume} {111}},\ \bibinfo {pages} {106401} (\bibinfo {year}
  {2013})}\BibitemShut {NoStop}%
\bibitem [{\citenamefont {Dobson}(1991)}]{DOBSON_JCP94_4328}%
  \BibitemOpen
  \bibfield  {author} {\bibinfo {author} {\bibfnamefont {J.~F.}\ \bibnamefont
  {Dobson}},\ }\href@noop {} {\bibfield  {journal} {\bibinfo  {journal} {J.
  Chem. Phys.}\ }\textbf {\bibinfo {volume} {94}},\ \bibinfo {pages} {4328}
  (\bibinfo {year} {1991})}\BibitemShut {NoStop}%
\bibitem [{\citenamefont {Dobson}(1993)}]{DOBSON_JCP98_8870}%
  \BibitemOpen
  \bibfield  {author} {\bibinfo {author} {\bibfnamefont {J.~F.}\ \bibnamefont
  {Dobson}},\ }\href {\doibase 10.1063/1.464444} {\bibfield  {journal}
  {\bibinfo  {journal} {J. Chem. Phys.}\ }\textbf {\bibinfo {volume} {98}},\
  \bibinfo {pages} {8870} (\bibinfo {year} {1993})},\ \Eprint
  {http://arxiv.org/abs/https://doi.org/10.1063/1.464444}
  {https://doi.org/10.1063/1.464444} \BibitemShut {NoStop}%
\bibitem [{\citenamefont {Becke}(1996)}]{BECKE_CJC74_995}%
  \BibitemOpen
  \bibfield  {author} {\bibinfo {author} {\bibfnamefont {A.~D.}\ \bibnamefont
  {Becke}},\ }\href@noop {} {\bibfield  {journal} {\bibinfo  {journal} {Can. J.
  Chem.}\ }\textbf {\bibinfo {volume} {74}},\ \bibinfo {pages} {995} (\bibinfo
  {year} {1996})}\BibitemShut {NoStop}%
\bibitem [{\citenamefont {Bates}\ and\ \citenamefont
  {Furche}(2012)}]{BATES_JCP137_164105}%
  \BibitemOpen
  \bibfield  {author} {\bibinfo {author} {\bibfnamefont {J.~E.}\ \bibnamefont
  {Bates}}\ and\ \bibinfo {author} {\bibfnamefont {F.}~\bibnamefont {Furche}},\
  }\href@noop {} {\bibfield  {journal} {\bibinfo  {journal} {J. Chem. Phys.}\
  }\textbf {\bibinfo {volume} {137}},\ \bibinfo {pages} {164105} (\bibinfo
  {year} {2012})}\BibitemShut {NoStop}%
\bibitem [{\citenamefont {Sagvolden}\ \emph {et~al.}(2013)\citenamefont
  {Sagvolden}, \citenamefont {Ekstr{\"o}m},\ and\ \citenamefont
  {Tellgren}}]{SAGVOLDEN_MP111_1295}%
  \BibitemOpen
  \bibfield  {author} {\bibinfo {author} {\bibfnamefont {E.}~\bibnamefont
  {Sagvolden}}, \bibinfo {author} {\bibfnamefont {U.}~\bibnamefont
  {Ekstr{\"o}m}}, \ and\ \bibinfo {author} {\bibfnamefont {E.~I.}\ \bibnamefont
  {Tellgren}},\ }\href {\doibase 10.1080/00268976.2013.802383} {\bibfield
  {journal} {\bibinfo  {journal} {Mol. Phys.}\ }\textbf {\bibinfo {volume}
  {111}},\ \bibinfo {pages} {1295} (\bibinfo {year} {2013})}\BibitemShut
  {NoStop}%
\bibitem [{\citenamefont {Zhu}\ \emph {et~al.}(2014)\citenamefont {Zhu},
  \citenamefont {Zhang},\ and\ \citenamefont {Trickey}}]{ZHU_PRA90_022504}%
  \BibitemOpen
  \bibfield  {author} {\bibinfo {author} {\bibfnamefont {W.}~\bibnamefont
  {Zhu}}, \bibinfo {author} {\bibfnamefont {L.}~\bibnamefont {Zhang}}, \ and\
  \bibinfo {author} {\bibfnamefont {S.~B.}\ \bibnamefont {Trickey}},\ }\href
  {\doibase 10.1103/PhysRevA.90.022504} {\bibfield  {journal} {\bibinfo
  {journal} {Phys. Rev. A}\ }\textbf {\bibinfo {volume} {90}},\ \bibinfo
  {pages} {022504} (\bibinfo {year} {2014})}\BibitemShut {NoStop}%
\bibitem [{\citenamefont {Furness}\ \emph {et~al.}(2015)\citenamefont
  {Furness}, \citenamefont {Verbeke}, \citenamefont {Tellgren}, \citenamefont
  {Stopkowicz}, \citenamefont {Ekstr{\"o}m}, \citenamefont {Helgaker},\ and\
  \citenamefont {Teale}}]{FURNESS_JCTC11_4169}%
  \BibitemOpen
  \bibfield  {author} {\bibinfo {author} {\bibfnamefont {J.~W.}\ \bibnamefont
  {Furness}}, \bibinfo {author} {\bibfnamefont {J.}~\bibnamefont {Verbeke}},
  \bibinfo {author} {\bibfnamefont {E.~I.}\ \bibnamefont {Tellgren}}, \bibinfo
  {author} {\bibfnamefont {S.}~\bibnamefont {Stopkowicz}}, \bibinfo {author}
  {\bibfnamefont {U.}~\bibnamefont {Ekstr{\"o}m}}, \bibinfo {author}
  {\bibfnamefont {T.}~\bibnamefont {Helgaker}}, \ and\ \bibinfo {author}
  {\bibfnamefont {A.~M.}\ \bibnamefont {Teale}},\ }\href {\doibase
  10.1021/acs.jctc.5b00535} {\bibfield  {journal} {\bibinfo  {journal} {J.
  Chem. Theory Comput.}\ }\textbf {\bibinfo {volume} {11}},\ \bibinfo {pages}
  {4169} (\bibinfo {year} {2015})}\BibitemShut {NoStop}%
\bibitem [{\citenamefont {Reimann}\ \emph {et~al.}(2017)\citenamefont
  {Reimann}, \citenamefont {Borgoo}, \citenamefont {Tellgren}, \citenamefont
  {Teale},\ and\ \citenamefont {Helgaker}}]{REIMANN_JCTC13_4089}%
  \BibitemOpen
  \bibfield  {author} {\bibinfo {author} {\bibfnamefont {S.}~\bibnamefont
  {Reimann}}, \bibinfo {author} {\bibfnamefont {A.}~\bibnamefont {Borgoo}},
  \bibinfo {author} {\bibfnamefont {E.~I.}\ \bibnamefont {Tellgren}}, \bibinfo
  {author} {\bibfnamefont {A.~M.}\ \bibnamefont {Teale}}, \ and\ \bibinfo
  {author} {\bibfnamefont {T.}~\bibnamefont {Helgaker}},\ }\href@noop {}
  {\bibfield  {journal} {\bibinfo  {journal} {J. Chem. Theory Comput.}\
  }\textbf {\bibinfo {volume} {13}},\ \bibinfo {pages} {4089} (\bibinfo {year}
  {2017})}\BibitemShut {NoStop}%
\bibitem [{\citenamefont {Tao}\ and\ \citenamefont
  {Perdew}(2005)}]{TAO_PRL95_196403}%
  \BibitemOpen
  \bibfield  {author} {\bibinfo {author} {\bibfnamefont {J.}~\bibnamefont
  {Tao}}\ and\ \bibinfo {author} {\bibfnamefont {J.~P.}\ \bibnamefont
  {Perdew}},\ }\href {\doibase 10.1103/PhysRevLett.95.196403} {\bibfield
  {journal} {\bibinfo  {journal} {Phys. Rev. Lett.}\ }\textbf {\bibinfo
  {volume} {95}},\ \bibinfo {pages} {196403} (\bibinfo {year}
  {2005})}\BibitemShut {NoStop}%
\bibitem [{\citenamefont {Becke}(1983)}]{BECKE_IJQC23_1915}%
  \BibitemOpen
  \bibfield  {author} {\bibinfo {author} {\bibfnamefont {A.~D.}\ \bibnamefont
  {Becke}},\ }\href@noop {} {\bibfield  {journal} {\bibinfo  {journal} {Int. J.
  Quantum Chem.}\ }\textbf {\bibinfo {volume} {23}},\ \bibinfo {pages} {1915}
  (\bibinfo {year} {1983})}\BibitemShut {NoStop}%
\bibitem [{\citenamefont {Tellgren}\ \emph
  {et~al.}(2014{\natexlab{b}})\citenamefont {Tellgren}, \citenamefont {Kvaal},\
  and\ \citenamefont {Helgaker}}]{TELLGREN_PRA89_012515}%
  \BibitemOpen
  \bibfield  {author} {\bibinfo {author} {\bibfnamefont {E.~I.}\ \bibnamefont
  {Tellgren}}, \bibinfo {author} {\bibfnamefont {S.}~\bibnamefont {Kvaal}}, \
  and\ \bibinfo {author} {\bibfnamefont {T.}~\bibnamefont {Helgaker}},\
  }\href@noop {} {\bibfield  {journal} {\bibinfo  {journal} {Phys. Rev. A}\
  }\textbf {\bibinfo {volume} {89}},\ \bibinfo {pages} {012515} (\bibinfo
  {year} {2014}{\natexlab{b}})}\BibitemShut {NoStop}%
\bibitem [{\citenamefont {Lieb}\ and\ \citenamefont
  {Schrader}(2013)}]{LIEB_PRA88_032516}%
  \BibitemOpen
  \bibfield  {author} {\bibinfo {author} {\bibfnamefont {E.~H.}\ \bibnamefont
  {Lieb}}\ and\ \bibinfo {author} {\bibfnamefont {R.}~\bibnamefont
  {Schrader}},\ }\href {\doibase 10.1103/PhysRevA.88.032516} {\bibfield
  {journal} {\bibinfo  {journal} {Phys. Rev. A}\ }\textbf {\bibinfo {volume}
  {88}},\ \bibinfo {pages} {032516} (\bibinfo {year} {2013})}\BibitemShut
  {NoStop}%
\bibitem [{\citenamefont {Rudelson}\ and\ \citenamefont
  {Vershynin}(2007)}]{RUDELSON_JACM54_21}%
  \BibitemOpen
  \bibfield  {author} {\bibinfo {author} {\bibfnamefont {M.}~\bibnamefont
  {Rudelson}}\ and\ \bibinfo {author} {\bibfnamefont {R.}~\bibnamefont
  {Vershynin}},\ }\href {\doibase 10.1145/1255443.1255449} {\bibfield
  {journal} {\bibinfo  {journal} {J. ACM}\ }\textbf {\bibinfo {volume} {54}}
  (\bibinfo {year} {2007}),\ 10.1145/1255443.1255449}\BibitemShut {NoStop}%
\bibitem [{\citenamefont {Roy}\ and\ \citenamefont
  {Vetterli}(2007)}]{ROY_ESPC_2007}%
  \BibitemOpen
  \bibfield  {author} {\bibinfo {author} {\bibfnamefont {O.}~\bibnamefont
  {Roy}}\ and\ \bibinfo {author} {\bibfnamefont {M.}~\bibnamefont {Vetterli}},\
  }in\ \href@noop {} {\emph {\bibinfo {booktitle} {2007 15th European Signal
  Processing Conference}}}\ (\bibinfo {year} {2007})\ pp.\ \bibinfo {pages}
  {606--610}\BibitemShut {NoStop}%
\bibitem [{\citenamefont {Dunning~Jr.}(1989)}]{DUNNING_JCP90_1007}%
  \BibitemOpen
  \bibfield  {author} {\bibinfo {author} {\bibfnamefont {T.~H.}\ \bibnamefont
  {Dunning~Jr.}},\ }\href@noop {} {\bibfield  {journal} {\bibinfo  {journal}
  {J. Chem. Phys.}\ }\textbf {\bibinfo {volume} {90}},\ \bibinfo {pages} {1007}
  (\bibinfo {year} {1989})}\BibitemShut {NoStop}%
\bibitem [{\citenamefont {Woon}\ and\ \citenamefont
  {Dunning~Jr.}(1994)}]{WOON_JCP100_2975}%
  \BibitemOpen
  \bibfield  {author} {\bibinfo {author} {\bibfnamefont {D.~E.}\ \bibnamefont
  {Woon}}\ and\ \bibinfo {author} {\bibfnamefont {T.~H.}\ \bibnamefont
  {Dunning~Jr.}},\ }\href@noop {} {\bibfield  {journal} {\bibinfo  {journal}
  {J. Chem. Phys.}\ }\textbf {\bibinfo {volume} {100}},\ \bibinfo {pages}
  {2975} (\bibinfo {year} {1994})}\BibitemShut {NoStop}%
\bibitem [{\citenamefont {Kendall}\ \emph {et~al.}(1992)\citenamefont
  {Kendall}, \citenamefont {Dunning~Jr.},\ and\ \citenamefont
  {J.}}]{KENDALL_JCP96_6796}%
  \BibitemOpen
  \bibfield  {author} {\bibinfo {author} {\bibfnamefont {R.~A.}\ \bibnamefont
  {Kendall}}, \bibinfo {author} {\bibfnamefont {T.~H.}\ \bibnamefont
  {Dunning~Jr.}}, \ and\ \bibinfo {author} {\bibfnamefont {H.~R.}\ \bibnamefont
  {J.}},\ }\href@noop {} {\bibfield  {journal} {\bibinfo  {journal} {J. Chem.
  Phys.}\ }\textbf {\bibinfo {volume} {96}},\ \bibinfo {pages} {6796} (\bibinfo
  {year} {1992})}\BibitemShut {NoStop}%
\bibitem [{\citenamefont {London}(1937)}]{LONDON_JPR8_397}%
  \BibitemOpen
  \bibfield  {author} {\bibinfo {author} {\bibfnamefont {F.}~\bibnamefont
  {London}},\ }\href@noop {} {\bibfield  {journal} {\bibinfo  {journal} {J.
  Phys. Radium}\ }\textbf {\bibinfo {volume} {8}},\ \bibinfo {pages} {397}
  (\bibinfo {year} {1937})}\BibitemShut {NoStop}%
\bibitem [{\citenamefont {Tellgren}\ \emph {et~al.}(2008)\citenamefont
  {Tellgren}, \citenamefont {Soncini},\ and\ \citenamefont
  {Helgaker}}]{TELLGREN_JCP129_154114}%
  \BibitemOpen
  \bibfield  {author} {\bibinfo {author} {\bibfnamefont {E.~I.}\ \bibnamefont
  {Tellgren}}, \bibinfo {author} {\bibfnamefont {A.}~\bibnamefont {Soncini}}, \
  and\ \bibinfo {author} {\bibfnamefont {T.}~\bibnamefont {Helgaker}},\
  }\href@noop {} {\bibfield  {journal} {\bibinfo  {journal} {J. Chem. Phys.}\
  }\textbf {\bibinfo {volume} {129}},\ \bibinfo {eid} {154114} (\bibinfo {year}
  {2008})}\BibitemShut {NoStop}%
\bibitem [{Lon()}]{LondonProgram}%
  \BibitemOpen
  \href@noop {} {\enquote {\bibinfo {title} {{LONDON, a quantum-chemistry
  program for plane-wave/GTO hybrid basis sets and finite magnetic field
  calculations. By E. Tellgren (primary author), T. Helgaker, A. Soncini, K. K.
  Lange, A. M. Teale, U. Ekstr{\"o}m, S. Stopkowicz, J. H. Austad, and S. Sen.
  See londonprogram.org for more information.}}}\ }\BibitemShut {NoStop}%
\bibitem [{\citenamefont {Tellgren}\ \emph {et~al.}(2009)\citenamefont
  {Tellgren}, \citenamefont {Helgaker},\ and\ \citenamefont
  {Soncini}}]{TELLGREN_PCCP11_5489}%
  \BibitemOpen
  \bibfield  {author} {\bibinfo {author} {\bibfnamefont {E.~I.}\ \bibnamefont
  {Tellgren}}, \bibinfo {author} {\bibfnamefont {T.}~\bibnamefont {Helgaker}},
  \ and\ \bibinfo {author} {\bibfnamefont {A.}~\bibnamefont {Soncini}},\
  }\href@noop {} {\bibfield  {journal} {\bibinfo  {journal} {Phys. Chem. Chem.
  Phys.}\ }\textbf {\bibinfo {volume} {11}},\ \bibinfo {pages} {5489} (\bibinfo
  {year} {2009})}\BibitemShut {NoStop}%
\bibitem [{\citenamefont {Becke}\ and\ \citenamefont
  {Edgecombe}(1990)}]{BECKE_JCP92_5397}%
  \BibitemOpen
  \bibfield  {author} {\bibinfo {author} {\bibfnamefont {A.~D.}\ \bibnamefont
  {Becke}}\ and\ \bibinfo {author} {\bibfnamefont {K.~E.}\ \bibnamefont
  {Edgecombe}},\ }\href@noop {} {\bibfield  {journal} {\bibinfo  {journal} {J.
  Chem. Phys.}\ }\textbf {\bibinfo {volume} {92}},\ \bibinfo {pages} {5397}
  (\bibinfo {year} {1990})}\BibitemShut {NoStop}%
\bibitem [{\citenamefont {Silvi}\ and\ \citenamefont
  {Savin}(1994)}]{SILVI_N371_683}%
  \BibitemOpen
  \bibfield  {author} {\bibinfo {author} {\bibfnamefont {B.}~\bibnamefont
  {Silvi}}\ and\ \bibinfo {author} {\bibfnamefont {A.}~\bibnamefont {Savin}},\
  }\href@noop {} {\bibfield  {journal} {\bibinfo  {journal} {Nature}\ }\textbf
  {\bibinfo {volume} {371}},\ \bibinfo {pages} {683} (\bibinfo {year}
  {1994})}\BibitemShut {NoStop}%
\bibitem [{\citenamefont {Furness}\ \emph {et~al.}(2016)\citenamefont
  {Furness}, \citenamefont {Ekstr{\"o}m}, \citenamefont {Helgaker},\ and\
  \citenamefont {Teale}}]{FURNESS_MP114_1415}%
  \BibitemOpen
  \bibfield  {author} {\bibinfo {author} {\bibfnamefont {J.~W.}\ \bibnamefont
  {Furness}}, \bibinfo {author} {\bibfnamefont {U.}~\bibnamefont
  {Ekstr{\"o}m}}, \bibinfo {author} {\bibfnamefont {T.}~\bibnamefont
  {Helgaker}}, \ and\ \bibinfo {author} {\bibfnamefont {A.~M.}\ \bibnamefont
  {Teale}},\ }\href@noop {} {\bibfield  {journal} {\bibinfo  {journal} {Mol.
  Phys.}\ }\textbf {\bibinfo {volume} {114}},\ \bibinfo {pages} {1415}
  (\bibinfo {year} {2016})}\BibitemShut {NoStop}%
\bibitem [{\citenamefont {Tao}\ \emph {et~al.}(2008)\citenamefont {Tao},
  \citenamefont {Vignale},\ and\ \citenamefont {Tokatly}}]{TAO_PRL100_206405}%
  \BibitemOpen
  \bibfield  {author} {\bibinfo {author} {\bibfnamefont {J.}~\bibnamefont
  {Tao}}, \bibinfo {author} {\bibfnamefont {G.}~\bibnamefont {Vignale}}, \ and\
  \bibinfo {author} {\bibfnamefont {I.~V.}\ \bibnamefont {Tokatly}},\ }\href
  {\doibase 10.1103/PhysRevLett.100.206405} {\bibfield  {journal} {\bibinfo
  {journal} {Phys. Rev. Lett.}\ }\textbf {\bibinfo {volume} {100}},\ \bibinfo
  {pages} {206405} (\bibinfo {year} {2008})}\BibitemShut {NoStop}%
\bibitem [{\citenamefont {Tao}\ \emph {et~al.}(2015)\citenamefont {Tao},
  \citenamefont {Liu}, \citenamefont {Zheng},\ and\ \citenamefont
  {Rappe}}]{TAO_PRB92_060401}%
  \BibitemOpen
  \bibfield  {author} {\bibinfo {author} {\bibfnamefont {J.}~\bibnamefont
  {Tao}}, \bibinfo {author} {\bibfnamefont {S.}~\bibnamefont {Liu}}, \bibinfo
  {author} {\bibfnamefont {F.}~\bibnamefont {Zheng}}, \ and\ \bibinfo {author}
  {\bibfnamefont {A.~M.}\ \bibnamefont {Rappe}},\ }\href {\doibase
  10.1103/PhysRevB.92.060401} {\bibfield  {journal} {\bibinfo  {journal} {Phys.
  Rev. B}\ }\textbf {\bibinfo {volume} {92}},\ \bibinfo {pages} {060401}
  (\bibinfo {year} {2015})}\BibitemShut {NoStop}%
\bibitem [{\citenamefont {Bader}(1980)}]{BADER_JCP73_2871}%
  \BibitemOpen
  \bibfield  {author} {\bibinfo {author} {\bibfnamefont {R.~F.~W.}\
  \bibnamefont {Bader}},\ }\href {\doibase 10.1063/1.440457} {\bibfield
  {journal} {\bibinfo  {journal} {J. Chem. Phys.}\ }\textbf {\bibinfo {volume}
  {73}},\ \bibinfo {pages} {2871} (\bibinfo {year} {1980})},\ \Eprint
  {http://arxiv.org/abs/https://doi.org/10.1063/1.440457}
  {https://doi.org/10.1063/1.440457} \BibitemShut {NoStop}%
\bibitem [{\citenamefont {Guevara-Garc{\'i}a}\ \emph
  {et~al.}(2011)\citenamefont {Guevara-Garc{\'i}a}, \citenamefont {Echegaray},
  \citenamefont {Toro-Labbe}, \citenamefont {Jenkins}, \citenamefont {Kirk},\
  and\ \citenamefont {Ayers}}]{GUEVARAGARCIA_JCP134_234106}%
  \BibitemOpen
  \bibfield  {author} {\bibinfo {author} {\bibfnamefont {A.}~\bibnamefont
  {Guevara-Garc{\'i}a}}, \bibinfo {author} {\bibfnamefont {E.}~\bibnamefont
  {Echegaray}}, \bibinfo {author} {\bibfnamefont {A.}~\bibnamefont
  {Toro-Labbe}}, \bibinfo {author} {\bibfnamefont {S.}~\bibnamefont {Jenkins}},
  \bibinfo {author} {\bibfnamefont {S.~R.}\ \bibnamefont {Kirk}}, \ and\
  \bibinfo {author} {\bibfnamefont {P.~W.}\ \bibnamefont {Ayers}},\ }\href
  {\doibase 10.1063/1.3599935} {\bibfield  {journal} {\bibinfo  {journal} {J.
  Chem. Phys.}\ }\textbf {\bibinfo {volume} {134}},\ \bibinfo {pages} {234106}
  (\bibinfo {year} {2011})},\ \Eprint
  {http://arxiv.org/abs/https://doi.org/10.1063/1.3599935}
  {https://doi.org/10.1063/1.3599935} \BibitemShut {NoStop}%
\bibitem [{\citenamefont {Entwistle}\ \emph {et~al.}(2016)\citenamefont
  {Entwistle}, \citenamefont {Hodgson}, \citenamefont {Wetherell},
  \citenamefont {Longstaff}, \citenamefont {Ramsden},\ and\ \citenamefont
  {Godby}}]{ENTWISTLE_PRB94_205134}%
  \BibitemOpen
  \bibfield  {author} {\bibinfo {author} {\bibfnamefont {M.~T.}\ \bibnamefont
  {Entwistle}}, \bibinfo {author} {\bibfnamefont {M.~J.~P.}\ \bibnamefont
  {Hodgson}}, \bibinfo {author} {\bibfnamefont {J.}~\bibnamefont {Wetherell}},
  \bibinfo {author} {\bibfnamefont {B.}~\bibnamefont {Longstaff}}, \bibinfo
  {author} {\bibfnamefont {J.~D.}\ \bibnamefont {Ramsden}}, \ and\ \bibinfo
  {author} {\bibfnamefont {R.~W.}\ \bibnamefont {Godby}},\ }\href {\doibase
  10.1103/PhysRevB.94.205134} {\bibfield  {journal} {\bibinfo  {journal} {Phys.
  Rev. B}\ }\textbf {\bibinfo {volume} {94}},\ \bibinfo {pages} {205134}
  (\bibinfo {year} {2016})}\BibitemShut {NoStop}%
\bibitem [{\citenamefont {Sun}\ \emph {et~al.}(2016{\natexlab{a}})\citenamefont
  {Sun}, \citenamefont {Perdew}, \citenamefont {Yang},\ and\ \citenamefont
  {Peng}}]{SUN_JCP144_19101}%
  \BibitemOpen
  \bibfield  {author} {\bibinfo {author} {\bibfnamefont {J.}~\bibnamefont
  {Sun}}, \bibinfo {author} {\bibfnamefont {J.~P.}\ \bibnamefont {Perdew}},
  \bibinfo {author} {\bibfnamefont {Z.}~\bibnamefont {Yang}}, \ and\ \bibinfo
  {author} {\bibfnamefont {H.}~\bibnamefont {Peng}},\ }\href {\doibase
  10.1063/1.4950845} {\bibfield  {journal} {\bibinfo  {journal} {J. Chem.
  Phys.}\ }\textbf {\bibinfo {volume} {144}},\ \bibinfo {pages} {191101}
  (\bibinfo {year} {2016}{\natexlab{a}})},\ \Eprint
  {http://arxiv.org/abs/https://doi.org/10.1063/1.4950845}
  {https://doi.org/10.1063/1.4950845} \BibitemShut {NoStop}%
\bibitem [{\citenamefont {Sun}\ \emph {et~al.}(2016{\natexlab{b}})\citenamefont
  {Sun}, \citenamefont {Perdew}, \citenamefont {Yang},\ and\ \citenamefont
  {Peng}}]{SUN_JCP145_019902}%
  \BibitemOpen
  \bibfield  {author} {\bibinfo {author} {\bibfnamefont {J.}~\bibnamefont
  {Sun}}, \bibinfo {author} {\bibfnamefont {J.~P.}\ \bibnamefont {Perdew}},
  \bibinfo {author} {\bibfnamefont {Z.}~\bibnamefont {Yang}}, \ and\ \bibinfo
  {author} {\bibfnamefont {H.}~\bibnamefont {Peng}},\ }\href {\doibase
  10.1063/1.4955444} {\bibfield  {journal} {\bibinfo  {journal} {J. Chem.
  Phys.}\ }\textbf {\bibinfo {volume} {145}},\ \bibinfo {pages} {019902}
  (\bibinfo {year} {2016}{\natexlab{b}})}\BibitemShut {NoStop}%
\end{thebibliography}
\end{document}